\documentclass[usenatbib]{mnras}
\usepackage{natbib}
\usepackage{apjfonts}
\usepackage{amsmath}
\usepackage[table]{xcolor}

\usepackage{epsfig}
\usepackage{graphicx}
\usepackage{amssymb}
\usepackage{aas_macros}
\usepackage{color}

\newcommand{\kms}{\,{\rm km \, s^{-1}}}
\newcommand{\pc}{\,{\rm pc}}
\newcommand{\kpc}{\,{\rm kpc}}
\newcommand{\msol}{\,M_\odot}

\newcommand{\bb}{\mathbfit }
\newcommand{\oversim}[2]{\protect{\mbox{\lower0.5ex\vbox{%
   \baselineskip=0pt\lineskip=0.2ex
   \ialign{$\mathsurround=0pt #1\hfil##\hfil$\crcr#2\crcr\sim\crcr}}}}} 
\catcode`\"=\active\let"=\"  

\def\3{{\ss} }

\def\c12{{1\over 2}}   
\def\erf{{\rm erf}}   
   
\def\d{{\rm d}}   
   
\def\plusplus{\raise 0.3ex\hbox{${\scriptstyle ++}$}{}}

\def\and{{{\rm M}31}}

\def\gyr{\,{\rm Gyr}}
\def\myr{\,{\rm Myr}}

\bibliography{biblio}

\usepackage{array}
\newcolumntype{L}[1]{>{\raggedright\let\newline\\\arraybackslash\hspace{0pt}}m{#1}}
\newcolumntype{C}[1]{>{\centering\let\newline\\\arraybackslash\hspace{0pt}}m{#1}}
\newcolumntype{R}[1]{>{\raggedleft\let\newline\\\arraybackslash\hspace{0pt}}m{#1}}

\begin{document}
\title[Ultra-wide binaries in tidal streams]{Creation/destruction of ultra-wide binaries in tidal streams}

\author[Jorge Pe\~{n}arrubia]{Jorge Pe\~{n}arrubia$^{1,2}$\thanks{jorpega@roe.ac.uk}\\
  $^1$Institute for Astronomy, University of Edinburgh, Royal Observatory, Blackford Hill, Edinburgh EH9 3HJ, UK\\
  $^2$Centre for Statistics, University of Edinburgh, School of Mathematics, Edinburgh EH9 3FD, UK
}
\maketitle  

\begin{abstract}
  This paper uses statistical and $N$-body methods to explore a new mechanism to form binary stars with extremely large separations ($\gtrsim 0.1\pc$), whose origin is poorly understood. Here, ultra-wide binaries arise via chance entrapment of unrelated stars in tidal streams of disrupting clusters. It is shown that (i) the formation of ultra-wide binaries is not limited to the lifetime of a cluster, but continues after the progenitor is fully disrupted, (ii) the formation rate is proportional to the {\it local} phase-space density of the tidal tails, (iii) the semimajor axis distribution scales as $p(a)\d a\sim a^{1/2}\d a$ at $a\ll D$, where $D$ is the mean interstellar distance, and (vi) the eccentricity distribution is close to thermal, $p(e)\d e= 2 e\d e$.
  Owing to their low binding energies, ultra-wide binaries can be disrupted by both the smooth tidal field and passing substructures. The time-scale on which tidal fluctuations dominate over the mean field is inversely proportional to the local density of compact substructures.
  Monte-Carlo experiments show that binaries subject to tidal evaporation follow $p(a)\d a\sim a^{-1}\d a$ at $a\gtrsim a_{\rm peak}$, known as \"Opik's law, with a peak semi-major axis that contracts with time as $a_{\rm peak}\sim t^{-3/4}$. In contrast, a smooth Galactic potential introduces a sharp truncation at the tidal radius, $p(a)\sim 0$ at $a\gtrsim r_t$.
 The scaling relations of young clusters suggest that most ultra-wide binaries arise from the disruption of low-mass systems. Streams of globular clusters may be the birthplace of hundreds of ultra-wide binaries, making them ideal laboratories to probe clumpiness in the Galactic halo.
\end{abstract}   


\section{Introduction}\label{sec:intro}
For many decades, astronomers have puzzled about the existence of ultra-wide binary stars with separations exceeding $\gtrsim 0.1\pc$ ($\approx 2\times 10^4\,{\rm AU}$).
To date, several fundamental questions remain poorly understood: where do these system form? what formation mechanism can explain their extreme separations? how can they possibly survive in a clumpy Galactic environment?

Given that the majority of field stars are born in star clusters (Lada \& Lada 2003), one may wonder whether these systems are the birth place of wide binaries. This seems unlikely, given that ultra-wide binaries are so widely separated that they hardly fit in them. Indeed, the number density of stars in young clusters typically exceeds $n\gtrsim 10^2\pc^{-3}$, which corresponds to a mean interstellar distance $D= (2\pi n)^{-1/3}\lesssim 0.1\pc$. On the other hand, isolated molecular cloud cores are even smaller and denser, with typical sizes $r\lesssim 0.1\pc$ (Ward-Thompson et al. 2007). 
Under these conditions, it is difficult to see how stellar pairs with separations $\gtrsim 0.1\pc$ can form, let alone survive, in a star-forming environment (see e.g. Scally, Clarke \& McCaughrean 1999; Parker et al. 2009; Elliott \& Bayo 2016; Deacon \& Kraus 2020).
The obvious alternative is formation via random {\it entrapment} of unrelated stars in the field. However, such occurrences are very unlikely given the low probability of close dynamical interactions (Makarov 2012).

Several mechanisms have been proposed in an attempt to circumnavigate the above issues:
(1) Kouwenhoven et al. (2010, hereafter K10) and Moeckel\& Bate (2010) show that wide binaries can form in star-forming regions during an early expansionary phase in which a large fraction of the natal gas cloud is expelled by stellar feedback. Using direct $N$-body simulations of clusters with super-virial and fractal initial conditions, K10 find bound pairs with a bimodal semimajor-axis distribution: a tight, high-energy peak associated with hard binaries that form at core collapse, and another at much larger separations that arises from random pairs that become self-gravitating in the waning tides of the expanding cluster.
(2) Moeckel \& Clarke (2011) explore a complementary formation channel in which soft binaries are created during the (relatively slow) expansion of a cluster driven by collisional relaxation. Within a cluster the population of wide binaries is close to statistical balance as pairs are continuously perturbed into and out of bound configurations. However, this balance breaks during the expansion of a cluster and the consequent lowering of the stellar density.
As the cluster expands the tidal field wanes and stellar pairs effectively freeze out of the creation-destruction cycle, thus becoming a population of self-gravitating binaries. However,
the formation efficiency is low: typicall 1 wide binary per cluster survives independently of the cluster initial conditions.
(3) Reipurth \& Mikkola (2012) show that wide binaries can also form via three-body interactions. In this case the third star acts as the energy sink, and is generally ejected with a large velocity, while the remainder pair tends to move on an eccentric orbit.
The {\it unfolding} of unstable triples into pairs happens on very short time-scales $\sim 1$--$100 \myr$, and generates about 2\% of bound pairs in a cluster. (4) Tokovinin (2017) explores the formation of wide binaries from adjacent stellar cores that move slowly relative to each other within a star-forming region. This mechanism matches the fraction and the separation distribution observed in young moving groups with ages between 10 and $100\myr$.

In contrast to star forming regions, the abundance of wide binaries in moving groups and stellar associations is remarkably high, with a fraction of stars in pairs that reaches $20$--$40\%$ of the total (e.g. Joncour et al. 2017), and a semimajor axis distribution derived from de-projecting the observed separation function that typically follows \"Opik (1924)'s law, $p(a)\d a\sim a^{-1}\d a$ (Kouwenhoven et al. 2007; Kraus \& Hillenbrand 2008, 2009). 
Binary stars with large separations are considerably more sparse in the field (e.g. Chanam\'e \& Gould 2004; Makarov et al. 2008). Fortunately, the advent of the Gaia mission (Gaia Collaboration et al. 2016, 2019) has dramatically expanded the sample of wide binaries detected with precise parallaxes and proper motions, which has led to the discovery of an excess of co-moving pairs with separations greater than 1 pc (Oh et al. 2017, Oelkers et al. 2017; Igoshev \& Perets 2019), although it remains unclear how many of those systems are formally bound or a result of chance alignments (e.g. Andrews et al. 2017).  
Observations of wide binaries in the field show that the separation function is well described by a triple power-law. At small separations, $s<0.01\pc$ the distribution of wide binaries follows \"Opik's law, $p(s)\d s\sim s^{-1}\d s$ (Andrews et al. 2017). At intermediate populations $0.01 \lesssim s/\pc\lesssim 0.1$, binary populations show a steeper separation function, $p(s)\d s\sim s^{-1.5}\d s$ (El-Badry \& Rix 2018), whereas on scales of ultra-wide binaries, $s>0.1\pc$, the separation function falls off more steeply. The exact shape of the distribution of ultra-wide binaries depends on whether these objects move with disc or halo orbits. Tian et al. (2020) analyzed $\sim 8200$ binaries in Gaia DR2 with separations $0.01<s/\pc<1$ and compared the properties of pairs orbiting in the MW disc and the stellar halo.  Intriguingly, ultra-wide binaries are more strongly suppressed in the halo, $p(s)\d s\sim s^{-2.5}\d s$, than in the disc, $p(s)\d s\sim s^{-2}\d s$, which is contrary to what might be expected if the steepening were due to gravitational interactions with molecular clouds or stars. Using El-Badry \& Rix (2018) catalogue, which compiles $\sim 55,000$ high-confidence binaries in Gaia DR2 with separations $<0.1\pc$, 
Tokovini (2020) finds that the eccentricity distribution of these systems is most likely `thermal', $p(e)\d e\approx 2 e\d e$, which is expected when orbital energies follow a Boltzmann function (Jeans 1928; Heggie 1975).
Independent clues on the formation of wide binaries can also be gathered from the relative composition between binary members. Andrews et al. (2019) find that bound pairs in the MW disc have very similar chemistry (typically within 0.1 dex), whereas stars with different origins show typical abundance differences of 0.3--0.4 dex. 
In addition, the mass ratios of wide pairs also put constraints on their formation. Recently, El-Badry et al. (2019) measure a significant excess of equal-mass binaries out to separations as large as $0.1\pc$, which is difficult to reconcile with models where `wide twins' form via core fragmentation, suggesting instead that they formed at closer separations and were subsequently widened by dynamical interactions in their birth environments.

Ultra-wide binaries are extremely fragile systems that can be easily unbound by chance encounters with compact substructures (Heggie 1975; Hills 1975; Weinberg et al. 1987) as well as by the Galactic mean field  (Heisler \& Tremaine 1986; Jiang \& Tremaine 2010). Furthermore, either one or both binary members may evolve off main sequence, inducing rapid mass loss and subsequent unbinding (e.g. Johnston et al. 2012; El-Badry \& Rix 2018); or dynamical evolution may also lead to progressive widening and dissolution within unstable triples (Reipurth, \& Mikkola 2012).
Clearly, the field population of wide binaries must be seen as a mixture of differently processed initial populations (Goodwin 2010), which introduces significant uncertainties in the dynamical modelling of these objects. Notwithstanding the above difficulties, the extreme sensitivity of wide binaries to the local dynamical environment has been used to (i) place limits on the density of MACHOs and other unseen material in the Galactic disc and halo (e.g. Bahcall, Hut \& Tremaine 1985; Yoo et al. 2004; Quinn et al. 2009), (ii) constrain the merger history of the Galaxy (e.g. Allen, Poveda \& Hern\'andez-Alc\'antara 2007), (iii) test Newtonian gravity at low accelerations (Jim\'enez et al. 2014; Scarpa et al. 2017; Banik \& Zhao 2018; El-Badry 2019; Pittordis \& Sutherland 2019; Hern\'andez et al. 2019), (iv) test the dark matter hypothesis in dwarf spheroidal galaxies (e.g. Hernandez \& Lee 2008; Pe\~narrubia et al. 2016), and (v) probe the existence of dark subhaloes devoid of stars predicted by cold dark matter models (Pe\~narrubia et al. 2010).

Most theoretical studies use stochastic methods to follow the dynamical evolution of wide binaries subject to repeated encounters with pointlike field objects. In these techniques, the collective effect of a fluctuating tidal field is modelled by coefficients that describe drift and diffusion in orbital energies (e.g. Chandrasekhar 1944; King 1977; Retterer \& King 1982; Weinberg et al. 1987). This approach has well-known shortcomings 
\begin{itemize}
\item Due to the singular force induced by point-masses, the distribution of nearby particles needs to be truncated at some arbitrarily-small radius, which leads to diffusion coefficients that are proportional to an ill-defined Coulomb logarithm (Chandrasekhar 1941a; Bar-Or et al. 2013; Pe\~narrubia 2019b).
\item Fokker-Planck equations do not behave well when applied to loosely-bound systems, $E\sim 0$, a region of energy space dubbed the `fringe' by Spitzer \& Shapiro (1972). 
  Direct-force experiments show that diffusion equations underestimate the `accelerated' unbinding of particles in the fringe (see Fig. 6 of Pe\~narrubia 2019a).
\item Binary stars are assumed to disappear instantaneously as soon as they become unbound.
  \item The smooth component of the Galactic tidal field is ignored.   
\end{itemize}
Jiang \& Tremaine (2010) inspected the last three issues with the aid of numerical models that solve self-consistently the equations of motions of binary stars in a smooth Galactic potential, adding random velocity `kicks' to mimic the effect of passing stars. The results show that (i) formally disrupted binaries can be brought back to an energetically-bound configuration during chance encounters with passing stars, and (ii) unbound pairs drift apart slowly in the Galaxy potential, which introduces significant correlations in the positions and velocities of disc stars on small scales.

This contribution has two chief goals: first, we show that the tidal tails of disrupting clusters are a natural birthplace for ultra-wide binaries. In contrast to other scenarios, the formation in tidal streams is not limited to an early expansionary phase of a cluster, but extends over its entire lifetime and continues long after it has been fully disrupted by the Galactic tidal field.
Second, we use a new stochastic technique presented in Pe\~narrubia (2019a,b; henceforth P19a,b) to follow the dynamical evolution of wide binaries in a clumpy environment, which extends the analysis of Chandrasekhar (1941b) to a population of {\it extended substructures} in dynamical equilibrium within the host galaxy. By considering compact objects with a vanishing (but non-zero) size, we avoid unnecessary ad-hoc truncations at strong forces (Pe\~narrubia 2018). As Jiang \& Tremaine (2010), we apply a Monte-Carlo method that injects random velocity impulses at individual time-steps of the binary orbit integration. This technique has been tested against direct-force experiments in previous contributions (see \S5 of P19a and \S3 of P19b), and is able to accurately reproduce the unbinding of binaries from the fringe, a process known as `tidal evaporation'.

 The paper is arranged as follows: Section~\ref{sec:form} explores the formation of ultra-wide binaries in the tidal debris of stellar clusters,  while Section~\ref{sec:disrup} analyzes the dynamical evolution of these objects in a clumpy Galactic potential. 
 Section~\ref{sec:model} describes the statistics of random pairs in a uniform, isotropic background with a Maxwellian velocity distribution. We demonstrate that the formation of bound stellar pairs is directly proportional to the local phase-space density of stream stars, $Q=\rho/\sigma^3$, and show that the distance distribution ultra-wide binaries scales as $p(a)\sim a^{1/2}$ for $a\ll D$, where $D$ is the average interstellar distance.
 Section~\ref{sec:nbody} uses $N$-body models to identify and characterize bound pairs in the tidal streams of clusters moving on circular orbits.
Sections~\ref{sec:smooth} and~\ref{sec:clumpy} analyze the disruption of ultra-wide binaries by the host galaxy potential and by passing substructures, respectively. 
Section~\ref{sec:MC} presents Monte-Carlo $N$-body experiments that follow the dynamical evolution of the bound pairs identified in \S\ref{sec:nbody} in a Milky Way-like galaxy with and without substructures.
Section~\ref{sec:discuss} summarizes the main limitations of our analysis, and discusses follow-up applications. A brief summary of our results is presented in Section~\ref{sec:summary}.

\section{Formation of wide binaries in tidal streams}\label{sec:form}
\subsection{Random pair statistics}\label{sec:model}
Consider a self-gravitating stellar system orbiting around a massive host galaxy and losing mass to tides. The escape process occurs through the Lagrange points L1 and L2 (e.g. Daniel et al. 2017), which leads to the formation of two tidal tails. The one associated with L1 is more gravitationally bound and has a lower angular momentum than the progenitor cluster, thus leading the motion of the system. The other has lower energy and higher angular momentum and trails it (see Fig.~9 of Pe\~narrubia 2006 for illustration).

Let us assume that the initial mass of the cluster is $M_c(t=0)$, and that the mass of single stars is on average $m_\star=0.5\msol$ (e.g. Kroupa 2002). The initial number of stars contained in the cluster therefore is $N_c=M_c(t=0)/m_\star$. Of those, $N_{\rm unb}(t)$ have been tidally stripped at the time $t$, hence the bound mass fraction can be simply written as $f_{\rm unb}(t)=1-M_c(t)/M_c(t=0)=N_{\rm unb}(t)/N_c$.
For reasons that will become apparent below, it is useful to define the mass-loss rate of a cluster as the fraction of stars that become unbound in the time interval $t,t+\Delta t$ 
\begin{align}\label{eq:rateunb}
\mathcal{R}_{\rm unb}(t)\equiv \frac{1}{N_c}\frac{N_{\rm unb}(t+\Delta t)-N_{\rm unb}(t)}{\Delta t}.
\end{align}

The distribution of unbound stars in the host potential is characterized by the number density $n(\bb R,t)$, which is normalized such that $N_{\rm unb}(t)=\int\d^3 R\,n(\bb R,t)$. Here, we use a capital vector ${\mathbfit R}$ to denote positions measured from the host galaxy centre. Our chief assumption is that two equal-mass stars with a combined mass $m_b=2m_\star$ become a gravitationally-bound {\it pair} if their relative distance ($r)$ and velocity ($v$) lead to a negative specific energy $E=v^2/2-Gm_b/r<0$. Notice that this simple criterium for binary formation neglects the presence of an external tidal field\footnote{Other authors use a definition of ``bound pair'' that relies on the Jacobi energy $E_J=E+\Phi_c<0$, where $\Phi_c$ is the centrifugal potential arising in a non-inertial rotating frame (e.g. Jiang \& Tremaine 2010). Appendix~A shows that our conclusions do not change if Jacobi energies are used instead of self-gravitating energies. }. 
  We will inspect this issue in Section~\ref{sec:disrup}.

 On very small scales, $a\lesssim |\nabla n/n|_{\bb R}^{-1}$, the number density can be assumed to be approximately constant, $n({\bb R}+{\bf r})\approx n({\bb R})=n$, which is usually known as the {\it local approximation}. Here, the probability\footnote{To simplify our notation, in this paper $p(x)$ denotes the probability to find a particle in the interval $x$,$x+\d x$, where $x$ the dimension of interest.} of finding the closest star at a relative position $\bb r$ can be calculated as (e.g. Chavanis 2009)
\begin{align}\label{eq:pr_closest}
 p(\bb r)\d^3 r \sim \exp\big(-\frac{4}{3}\pi r^3 n\big )4\pi r^2 n \d r.
\end{align} 
It is straightforward to show that the function $p(\bb r)$ peaks at an `inter-stellar' distance $D\equiv (2\pi n)^{-1/3}$, which corresponds to the average distance between stars. In what follows, we will assume that the distance between bound pairs is much smaller than the average separation between stream particles, $a\lesssim D$, which implies $(4\pi/3) a^3\,n=(2/3)(a/D)^3\ll 1$. Thus, on scales $r\lesssim a$ the distribution of nearby stars~(\ref{eq:pr_closest}) becomes approximately homogeneous, $p(\bb r)\d^3 r\approx 4\pi r^2n\d r$. In addition, it is helpful to assume that the relative velocity distribution of nearby stars is Maxwellian, $p(\bb v)=(2\pi \sigma^2)^{-3/2}\exp[-v^2/(2\sigma^2)]$, where $\sigma=\sigma(\bb R,t)$ is the local, one-dimensional velocity dispersion of the tidal tails.

In the local and Maxwellian approximations, the number of energetically-bound stars that can be found within a volume $V=4\pi a^3/3$ centred at $\bb R$ at the time $t$ can be estimated from Equation~(\ref{eq:pr_closest}) as
\begin{align}\label{eq:DelN}
   N_b({\bb R},t)&=\int_V\d^3 r\, p(\bb r) \int_{E<0}\d^3v\,p(\bb v) \\\nonumber
  &=(4\pi)^2\frac{n}{(2\pi\sigma^2)^{3/2}}\int_0^a\d r\, r^2\int_0^{v_{e}(r)}\d v\,v^2 \exp\bigg(-\frac{v^2}{2\sigma^2}\bigg)\\\nonumber
  &=4\pi n \int_0^a\d r\, r^2\bigg\{\erf\bigg(\frac{v_e}{\sqrt{2}\sigma}\bigg)-\sqrt{\frac{2}{\pi}}\frac{v_e}{\sigma}\exp\bigg[-\frac{v_e^2}{2\sigma^2}\bigg]\bigg\},
\end{align}
where $v_e(r)=\sqrt{2 G m_b/r}$ is the escape speed of a binary system. For stellar binaries the escape speed is much smaller than the velocity dispersion of stream stars, and one can safely assume $v_e/\sigma\ll 1$. Hence, the integrand term within brackets can be approximately written as
$$\erf\bigg(\frac{x}{\sqrt{2}}\bigg)-\sqrt{\frac{2}{\pi}}\exp\bigg(-\frac{x^2}{2}\bigg)=\frac{1}{3}\sqrt{\frac{2}{\pi}}x^3 +\mathcal{O}(x^4).$$
At leading order, Equation~(\ref{eq:DelN}) becomes
\begin{align}\label{eq:DelN2}
   N_b({\bb R},t)&\approx \frac{4 \pi}{3} n \sqrt{\frac{2}{\pi}}\int_0^a\d r\, r^2 \bigg(\frac{v_e}{\sigma}\bigg)^3\\ \nonumber
  &=\frac{32\sqrt{\pi}}{9}(G m_b a)^{3/2}Q(\bb R,t),
\end{align} 
where $Q(\bb R,t)=n/\sigma^3$ is the {\it local phase-space density} of stream stars. Note that the precise value of the distance $a$ remains arbitrary in our derivation. This freedom arises from the construction of models in isolation, a point to which we return in Section~\ref{sec:disrup}.

The probability of finding a single star\footnote{Associations with triple or higher-order multiple stars are neglected in our analysis for simplicity.}
at a distance smaller than the average separation of stream members given by~(\ref{eq:pr_closest}) is $p(a)\sim (a/D)^2\ll 1$ for $a\lesssim D$.
Hence, the value of $N_b({\bb R},t)$ is expected to fluctuate strongly along tidal tails.
The analysis below simplifies considerably by introducing stream-averaged quantities. To do that, we multiply both sides of Equation~(\ref{eq:DelN2}) by $N_{\rm unb}^{-1} \int \d^3 R\sum_{i=1}^{N_{\rm unb}}\delta(\bb R-\bb R_i)$ and divide by the number of unbound stars. Integrating over volume yields a binary fraction
\begin{align}\label{eq:Nb}
f_b(t)\equiv\frac{N_b(t)}{N_{\rm unb}(t)}=\frac{32\sqrt{\pi}}{9}(G m_b a)^{3/2}\langle Q(t)\rangle,
\end{align}
where $N_b(t)=\int \d^3 R\sum_{i=1}^{N_{\rm unb}}\delta(\bb R-\bb R_i) N_b(\bb R,t)=\sum_{i=1}^{N_{\rm unb}}N_b(\bb R_i,t)$ is the total number of that stream stars that become a bound pair at a given time, and $\langle Q(t)\rangle=N_{\rm unb}^{-1}\sum_{i=1}^{N_{\rm unb}}Q(\bb R_i,t)$ is the {\it mean phase-space density} of the stream.

It is useful to define the {\it binary formation rate} as the fraction of stream members that become gravitationally bound to a neighbour star within the time interval $t,t+\Delta t$. From Equation~(\ref{eq:Nb})
\begin{align}\label{eq:ratef}
   \mathcal{R}_f(t)&\equiv\frac{f_b(t+\Delta t)-f_b(t)}{\Delta t}\\ \nonumber
  &=\frac{32\sqrt{\pi}}{9}(G m_b a)^{3/2}\frac{\langle Q(t+\Delta t)\rangle-\langle Q(t)\rangle}{\Delta t}.
\end{align}
Unfortunately, tidal streams exhibit non-trivial variations of phase-space density as a function of position and time, which greatly complicates any attempt to compute global formation rates analytically. Further physical insight into the binary formation process will be gained in Section~\ref{sec:nbody} with the aid of $N$-body experiments.

Crucially, the {\it distance distribution} of bound pairs can be derived from Equation~(\ref{eq:Nb}) as
\begin{align}\label{eq:fa}
  p(a,t)\equiv\frac{1}{N_{\rm unb}(t)}\frac{\d N_b(t)}{\d a}= p_0(t) \,a^{1/2},
\end{align}
where $p_0(t)=(16\sqrt{\pi}/3)(G m_b)^{3/2}\langle Q(t)\rangle$ is a normalization factor. Comparison between~(\ref{eq:DelN2}) and~(\ref{eq:Nb}) shows that $p(a,t)\propto a^{1/2}$ independently of position along the tidal tails and time. We will come back to this important result in \S\ref{sec:binary}.

 If one assumes that binary stars move on circular orbits, the energy distribution associated with~(\ref{eq:fa}) can be derived from the transformation $f(E)\,\d E=p(a)\d a$ with $a=Gm_b/(-2\,E)$. This returns
\begin{align}\label{eq:fE}
  f(E)\,\d E=f_0\,\bigg(\frac{Gm_b}{2}\bigg)^{3/2}\frac{\d E}{(-E)^{5/2}},
\end{align}
which vanishes in the limit $E\to -\infty$ ($a\to 0$), reflecting the low probability of forming tightly-bound pairs through random superposition of stream orbits. 

\subsection{$N$-body experiments}\label{sec:nbody}
This Section presents a number of $N$-body experiments that follow the tidal disruption of self-gravitating collisionless clusters in a Galactic potential and serve to illustrate the entrapment of particle pairs in tidal tails.

\subsubsection{Numerical set-up}\label{sec:setup}
Cluster models follow a Dehnen (1993) cored ($\gamma=0$) profile at $t=0$
\begin{align}\label{eq:rho}
  \rho_c(r)=\frac{3 M_c}{4\pi r_c^3}\frac{1}{(1+r/r_c)^4},
\end{align}
where $M_c$ and $r_c$ are the mass and scale-length, respectively.
Equilibrium $N$-body realizations of isotropic Dehnen spheres are generated via an Eddington (1916) inversion (see Errani \& Pe\~narrubia 2020 for details). Cluster models have initial masses $\log_{10}(M_c/M_\odot)=2.0, 2.5, 3.0, 3.5, 4.0, 4.5$, and a fixed scale-length $r_c=5\pc$. It will be shown below that the fraction of binary stars decreases with cluster mass. To remedy this issue and improve statistics, our cluster models have a varying number of $N$-body particles: $N_p= 10^5$ for $M_c<10^3\msol$; $N_p=2\times 10^5$ for $10^3<M_c/\msol\le 10^4\msol$, and $N_p=6\times 10^5$ for $M_c> 10^4\msol$.
For simplicity, clusters are injected on circular orbits in an analytical host potential that roughly matches the mass distribution of the Milky Way. The Galaxy model consists of a Hernquist (1990) bulge with a mass $M_b=2.3\times 10^{10}\msol$ and $c=1.2\kpc$, a Miyamoto \& Nagai (1970) disc with a mass $M_d=6.6\times 10^{10}\msol$, a scale length $a=8\kpc$ and scale height $b=0.3\kpc$, and a spherically-symmetric Navarro, Frenk \& White (1997) dark matter halo with a virial mass $M_h=10^{12}\msol$, a scale radius $R_s=21\kpc$ and a concentration $c_{\rm vir}=12.3$. To simplify the analysis, cluster orbits are confined in the disc plane ($z=0$).

\begin{figure}
\begin{center}
\includegraphics[width=80mm]{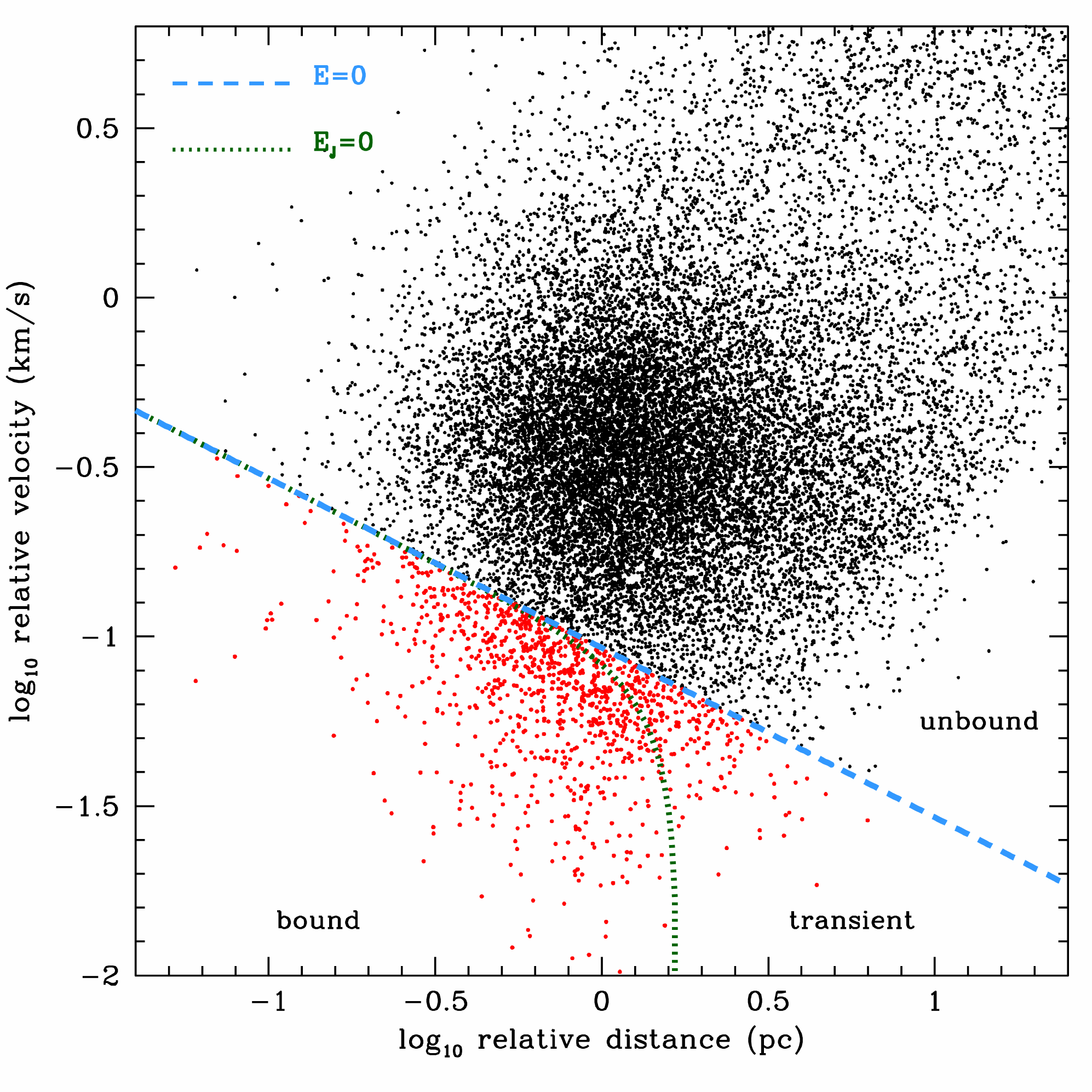}
\end{center}
\caption{Relative distance and velocity of the closest pair of 20000 particles randomly selected from the tidal debris of a cluster model with an initial mass $M_c=300\,M_\odot$, a scale length $r_c=5\pc$, moving on a circular orbit at a galactocentric radius $R=8\kpc$. Red particles denote gravitationally-bound pairs with a specific energy $E=v^2/2-Gm_b/r<0$ and a combined mass $m_b=1\msol$. For ease of reference, $E=0$ is marked with a blue-dashed line. The Jacobi energy $E_J=E+\Phi_c=0$ is marked with a green-dotted line (see text). Note that particle pairs with negative binding energies ($E<0$) and positive Jacobi energies ($E_J>0$) can be easily disrupted by the host tidal field and do not last long as self-gravitating objects (see Appendix~A). }
\label{fig:phase}
\end{figure}
Our cluster models lose mass to Galactic tides at a rate that depends on their density and orbital radius ($R$). Particles that become energetically unbound from the progenitor cluster are labelled members of the associated tidal stream.
To follow the process of tidal stripping we use {\sc superbox} (see Fellhauer et al. 2000), a highly-efficient particle-mesh algorithm that computes the gravitational potential by placing two co-moving grids centred at the densest region of a self-gravitating system. Each grid has $128^3$ cubic cells with sizes $\Delta x = 2r_c/128$ and $40 r_c /128$.
It should be stressed that {\sc superbox} is a collision-less code, which means that the results of our $N$-body experiments ignore the dynamical formation/disruption of binary stars that naturally occurs within stellar clusters during multiple-body encounters (e.g. Heggie 1975). This shortcoming is discussed in some detail in \S\ref{sec:discuss}.  
 The integration time is set to $t_{\rm now}=5\gyr$, which is representative of the age of stellar streams associated with globular clusters (e.g. Erkal et al. 2017; Malhan \& Ibata 2019). Snapshots are recorded at fixed time intervals separated by $\Delta t=0.05\gyr$. The time-step of the $N$-body models is set to $1/40{\rm th}$ of the dynamical time of the cluster, $t_{\rm dyn}=r_c^{3/2}/(G M_c)^{1/2}$, which is sufficiently short to guarantee dynamical equilibrium when models are run in isolation.

\subsubsection{Results}\label{sec:results}
Fig.~\ref{fig:phase} shows the relative velocity and distance of the closest neighbour of $20000$ stream particles randomly chosen from a model with an initial mass $M_c=300\msol$ and a scale-length $r_c=5\pc$ integrated for $t_{\rm now}=5\gyr$. Points are colour-coded in red if the phase-space locations of the particle pairs are sufficiently close as to yield a negative specific binding energy $E=v^2/2-Gm_b/r<0$, with $m_b=1\msol$. The fraction of particles under the line $E=0$ (blue-dashed line) is proportional to the mean phase-space density of stream particles. 
One can see by eye that the relative distance and velocity between 
random pairs peaks at $D\sim 1.3 \pc$ and $\langle v\rangle \sim 0.3\kms$, respectively, both appreciably larger than the typical separation and critical velocity of wide binaries, $r\lesssim 1\pc$ and $v_{\rm crit}<(2G m_b/r)^{1/2}=(2G\msol/\pc)^{1/2}\sim 0.1\kms$. This difference increases in streams associated with more massive clusters, for which both $D$ and $\langle v\rangle$ tend to have larger values. 
Fig.~\ref{fig:phase} reveals a few points of interest. First and foremost, it indicates that the formation of wide binaries in tidal tails are rare, low-probability events, as the vast majority of random pairs have self-gravitating energies $E>0$. Second, the number of bound pairs quickly drops at small ($r\lesssim 0.1\,D$) and large ($r\gtrsim 10\,D$) separations, with the majority having a separation comparable to the mean inter-stellar distance of the tidal tails ($r\sim D)$. Third, we find bound pairs with separations as large as $r\sim 7\pc$, a distance that surpasses the original scale-length of the cluster. However, many of these pairs have positive Jacobi energies $E_J=E+\Phi_c>0$, where $\Phi_c=(1/2)\Omega_g^2r^2$ is the centrifugal potential and $\Omega_g(R)=v_c(R)/R$ is the circular frequency of the pair's orbit about the host, which suggests that they can be easily disrupted by the underlying host potential and therefore have a transient nature. Section~\ref{sec:disrup} and Appendix~A inspect this important issue in depth.

\begin{figure}
\begin{center}
\includegraphics[width=86mm]{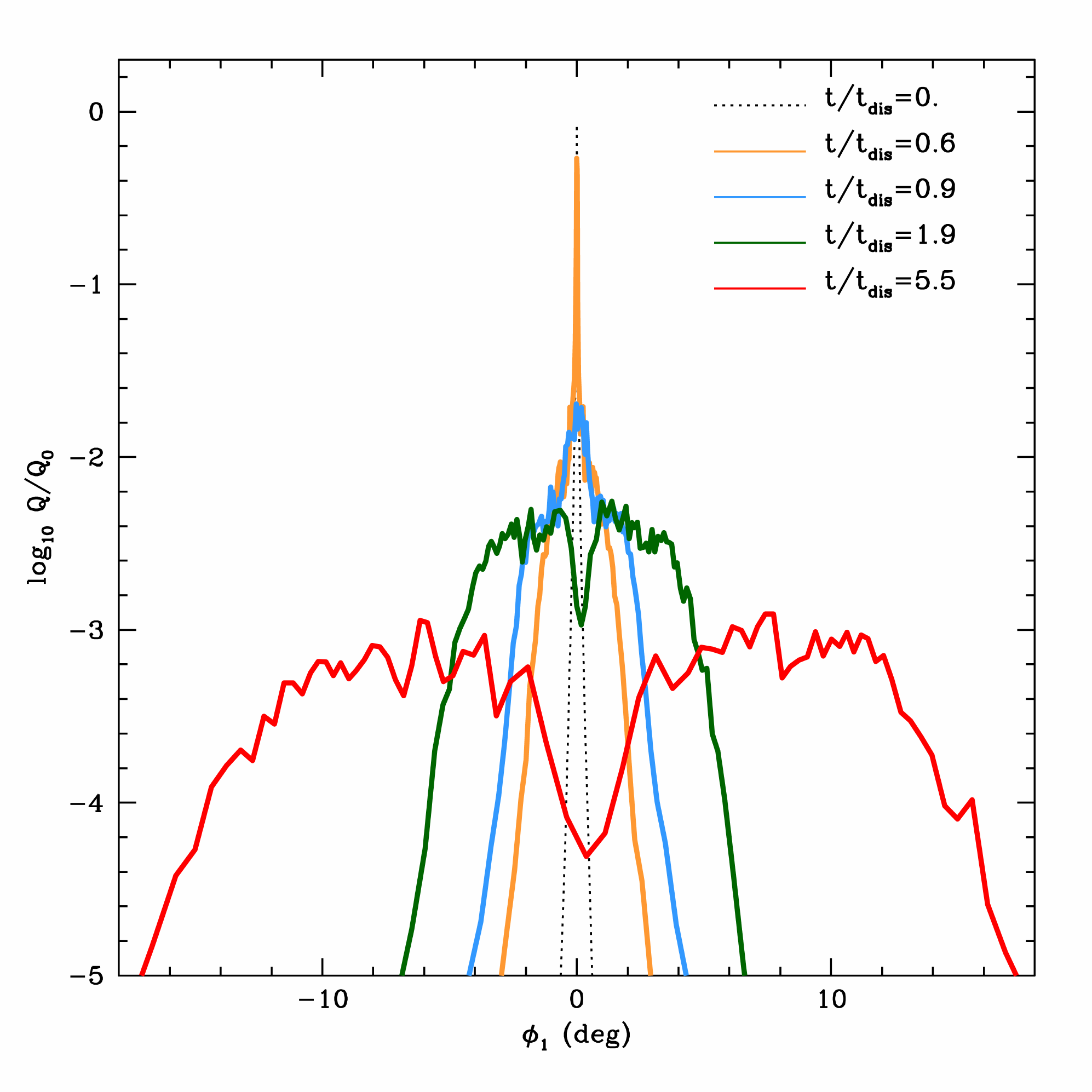}
\end{center}
\caption{Phase-space density of a stellar cluster losing mass to tides at different snapshots as a function of position angle along the stream ($\phi_1$). The cluster has an initial mass $M_c=300\,M_\odot$, a scale length $r_c=5\pc$ and moves on a circular orbit at a galactocentric radius $R=8\kpc$. Phase-space density is normalized by the cluster's central value, $Q_0$, Equation~(\ref{eq:Q0}). Time is measured in units of the disruption time, $t_{\rm dis}=0.9\gyr$. Note that unbound particles move away from the progenitor. Once tidal disruption is complete, a sharp drop of phase-space density is visible at the location of the (disrupted) cluster, $\phi_1=0$. }
\label{fig:Q}
\end{figure}

As demonstrated in Section~\ref{sec:model}, the probability that two stars form a bound system is proportional to the local phase-space density of stream particles. Unfortunately, clusters acted on by tides exhibit non-trivial variations of phase-space density in position and time, which adds considerable complexity to our analysis. As an illustration, Fig.~\ref{fig:Q} plots the phase-space density profile of a cluster model with an initial mass $M_c=300\msol$ and an orbital radius $R=8\kpc$ at different snapshots. Phase-space densities are measured in bins of position angle along the stream plane ($\phi_1$), with the progenitor cluster located at $\phi_1=0$ at all times. Here, the time is given in units of the cluster's disruption time, which corresponds to $t_{\rm dis}=0.9\gyr$ for this particular model. Dotted lines show the initial ($t=0$) phase-space density profile, which can be expressed analytically for Dehnen spheres as
\begin{align}\label{eq:Qc}
  Q(r)&=\frac{\rho_c(r)}{m_\star\sigma_r^3(r)}=\frac{Q_{0}}{(1+r/r_c)(1+6r/r_c)^{3/2}},
\end{align}
with a central value
\begin{align}\label{eq:Q0}
  Q_{0}= \frac{45}{\pi}\sqrt{\frac{15}{2}}\frac{1}{(G^{3}m_\star^2 M_cr_c^3)^{1/2}},
\end{align}
where $\sigma_r(r)$ is the radial velocity dispersion profile (eq. A3 of Dehnen 1993). One can easily see that Equation~(\ref{eq:Qc}) converges asymptotically to a maximum value $Q(r)\approx Q_{0}$  at small radii, $r\ll r_c$, whereas at large radii $r\gg r_c$ the phase-space density falls as $Q(r)\sim  r^{-5/2}$.

Comparison of the phase-space density profile at $t=0.6\,t_{\rm dis}$ with the initial profile given by Equation~(\ref{eq:Qc}) (dotted lines) shows that tidal mass loss lowers the density in the central regions of the cluster ($|\phi_1|\lesssim 1^\circ$). The presence of unbound material in the outskirts of the cluster manifests as a shallowing of the profile at $|\phi_1|\gtrsim 1^\circ$. At $t=0.9\,t_{\rm dis}$, shortly before tidal disruption is complete, the cluster's phase-space density has dropped two orders of magnitude below the initial central value, $Q(\phi_1=0)\sim 0.01\,Q_0$, whereas the mean phase-space density of the tidal tails is even lower, $Q(|\phi_1|>2^\circ)\lesssim 0.003\,Q_0$. After the cluster is fully disrupted ($t>t_{\rm dis}$), two salient features of Fig.~\ref{fig:Q} stand out: (i) a strong dip in the phase-space density appears at the location of the (disrupted) cluster ($\phi_1=0$), and (ii) the mean phase-space density decreases monotonically with time as tidally-stripped stars move progressively away from the progenitor system and of tidal tails phase-mix in the host galaxy potential.

\begin{figure}
\begin{center}
\includegraphics[width=84mm]{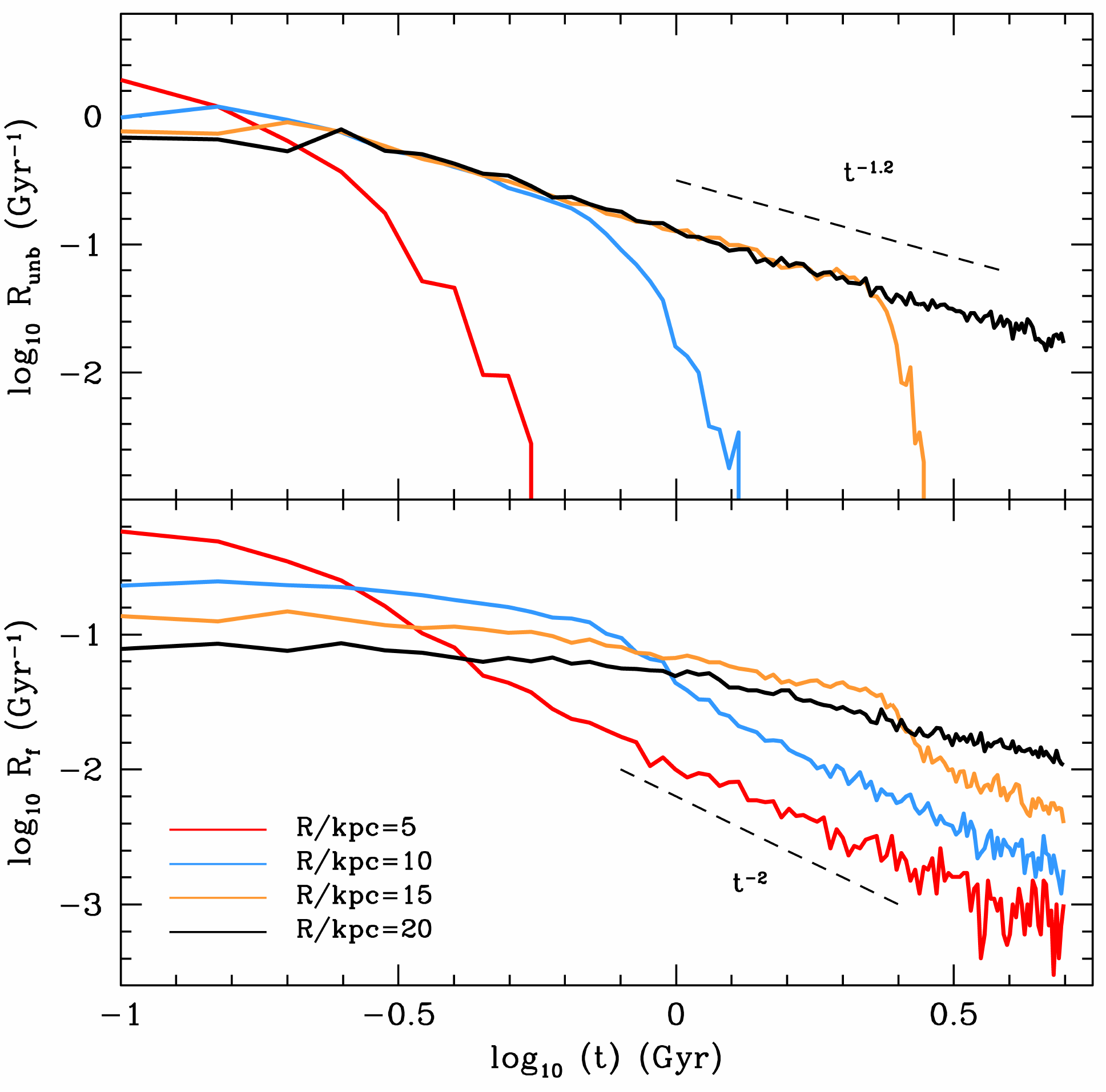}
\end{center}
\caption{{\it Upper panel}: mass-loss rate, Equation~(\ref{eq:rateunb}), of a cluster with an initial mass $M_c=300M_\odot$ and a scale-length $r_c=5\pc$ as a function of time. Cluster models move on circular orbits in a Milky Way-like disc plane at different galactocentric radii. Note the sharp drop of the mass-loss rate prior to full tidal disruption. {\it Lower panel}:  Formation rate, Equation~(\ref{eq:ratef}), of binaries with a separat ion $<1\pc$ as a function of time. At early times the formation rate if roughly constant, falling as $t^{-2}$ after the cluster is fully disrupted.}
\label{fig:rate_evol}
\end{figure}

The above results suggest that the average phase-space density of stellar streams depends on the mass loss history of the progenitor system.
To analyze this issue, we show in the upper panel of Fig.~\ref{fig:rate_evol} the mass-loss rate defined by Equation~(\ref{eq:rateunb}) of clusters with an initial mass $M_c=300\msol$ placed at different galactocentric radii. Notably, after a short time interval the evolution quickly becomes scale-free wherein the mass-loss rate scales as $\mathcal{R}_{\rm unb} \sim t^{-1.2}$. This power-law regime breaks at the time a cluster becomes fully disrupted, $t\approx t_{\rm dis}$, where a sudden drop of the mass-loss rate that marks the beginning of a run-away process that ends with the full unbinding of systems with cored density profiles (Errani \& Pe\~narrubia 2020).

In the lower panel of Fig.~\ref{fig:rate_evol} we plot the formation rate of bound ($E<0$) pairs with a relative distance $<1\pc$ as a function of time, Equation~(\ref{eq:ratef}). To identify stream particles that become bound we compute the separations and velocities of the closest 50 neighbours and sort them by specific energy. Particles with the lowest (negative) specific energy form a bound pair. To avoid duplication, we record their IDs and remove them from the list of particles that can become wide binaries at a later snapshot.
The evolution of the formation rates in Fig.~\ref{fig:rate_evol} exhibit two distinct regimes. At an early stage, $t\ll t_{\rm dis}$, binary form at an approximately constant rate, $\mathcal{R}_f\approx \mathcal{R}_{f,0}$. Models at larger galactocentric distances exhibit systematically lower values of $\mathcal{R}_{f,0}$. The disruption of the progenitor cluster leads to a pronounced change in the binary formation rate, which starts to fall as a power-law curve, $\mathcal{R}_f\sim t^{-2}$, after the cluster is fully dissolved. This behaviour can be modelled with a broken power-law
\begin{eqnarray}\label{eq:Rf0t0}
  \mathcal{R}_{f}(t)=
  \begin{cases}
     \mathcal{R}_{f,0}& ~~\, t< t_{\rm dis}\\
     \mathcal{R}_{f,0}\big(t/t_{\rm dis})^{-2} &~~\,t\ge t_{\rm dis},
    \end{cases}
\end{eqnarray}
which is uniquely defined by two parameters, a flat formation rate, $\mathcal{R}_{f,0}$, and the cluster's disruption time, $t_{\rm dis}$.
A closer look to the bottom panel of Fig.~\ref{fig:rate_evol} suggests that the two parameters are reciprocally related. In particular, clusters that are quickly destroyed by the Galactic tidal field tend to form bound stellar pairs at a higher rate and vice versa.
The {\it maximum} formation rate can computed from~(\ref{eq:ratef}) by replacing the phase-space density of stream stars by the cluster's central value at $t=0$, and noting that $\langle Q\rangle \ll Q_0$ (see Fig.~\ref{fig:Q}). Hence, combining Equations~(\ref{eq:ratef}) and~(\ref{eq:Q0}) and adopting $m_b=2m_\star$ yields
\begin{align}\label{eq:ratefmax}
  \mathcal{R}_{f,{\rm max}}(t)&\equiv \frac{32\sqrt{\pi}}{9}(G m_b a)^{3/2}\frac{Q_0}{\Delta t}\\ \nonumber
   &=320\sqrt{\frac{15}{\pi}}\bigg(\frac{m_\star }{M_c }\bigg)^{1/2}\bigg(\frac{a}{r_c}\bigg)^{3/2}\frac{1}{\Delta t}.
\end{align}

\begin{figure}
\begin{center}
\includegraphics[width=85mm]{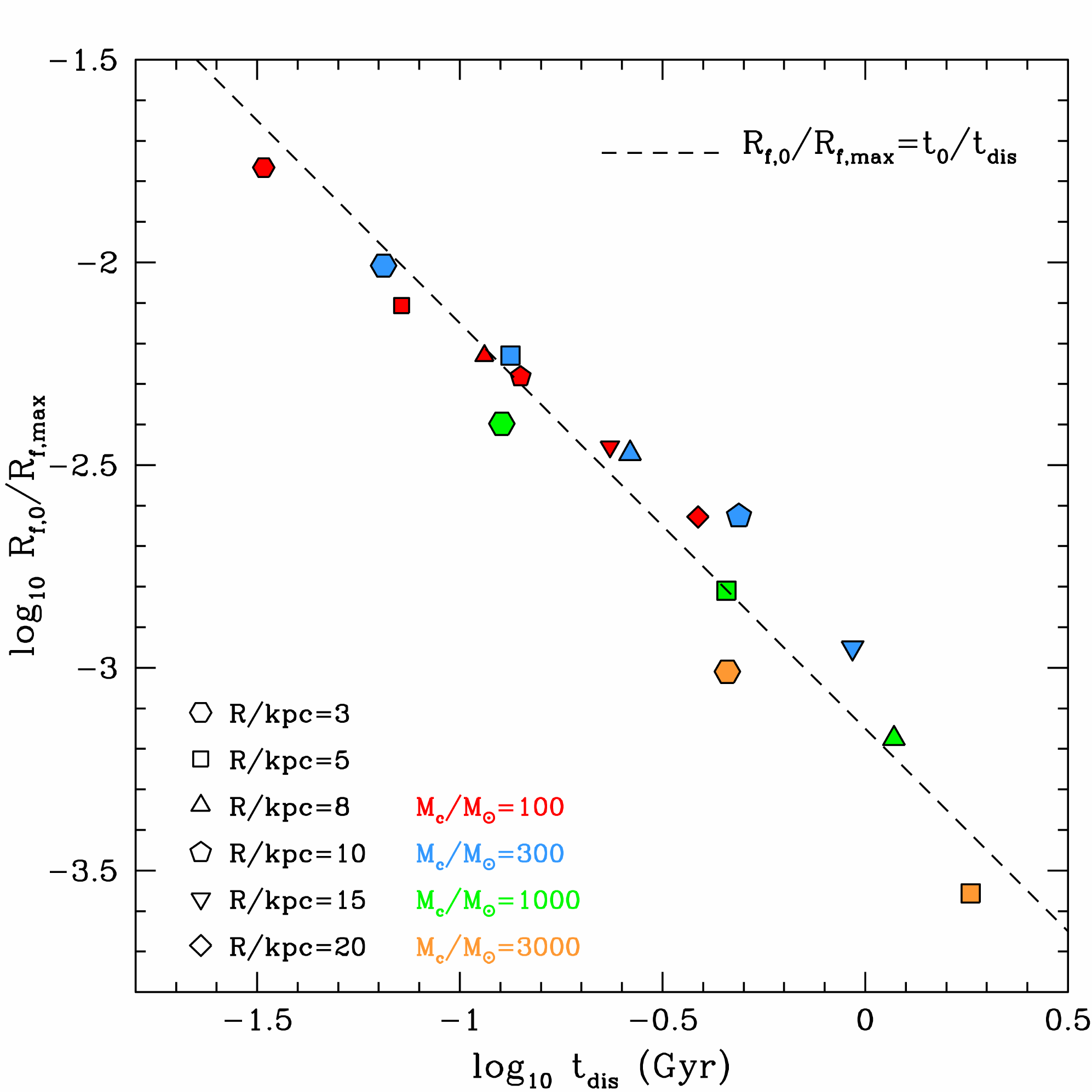}
\end{center}
\caption{Best-fit parameters $\mathcal{R}_{f,0}$ and $t_{\rm dis}$ of Equation~(\ref{eq:Rf0t0}) obtained from various $N$-body experiments covering a wide range of orbital radii ($R$) and cluster masses ($M_c$). All cluster models have an initial scale-length $r_c=5\pc$. A linear relation $\mathcal{R}_{f,0}/\mathcal{R}_{f,{\rm max}}=t_0/t_{\rm dis}$ with $t_0/\Delta t=0.028$ (black-dashed line) provides a reasonable description of the fitted parameters.}
\label{fig:rate_tdis}
\end{figure}

Interestingly, Fig.~\ref{fig:rate_tdis} shows that during the early stages of tidal stripping ($t\lesssim t_{\rm dis}$) the formation of bound pairs occurs at a rate that is inversely proportional to cluster's disruption time, i.e. $\mathcal{R}_{f,0}\sim t^{-1}_{\rm dis}$. In this plot, the values of 
$\mathcal{R}_{f,0}$ and $t_{\rm dis}$ are measured by fitting~(\ref{eq:Rf0t0}) to  the formation rates of cluster models with different masses and orbital radii. We find that the disruption time depends on (i) the pericentre of the orbit; (ii) the initial mass (and size) of the cluster; as well as (iii) the host potential. The individual impact of these parameters is highly degenerate, so that a particular modification in the survival of a cluster may be achieved by trading off the effect of one against another. A certain $t_{\rm dis}$, for example, may be achieved by a low-mass cluster of moderate pericentre, or by a massive one with smaller orbital radius.

The empirical relation shown in Fig.~\ref{fig:rate_tdis} can be roughly described by a linear function
\begin{align}\label{eq:linear}
  \frac{\mathcal{R}_{f,0}}{\mathcal{R}_{f,{\rm max}}}=\frac{t_0}{t_{\rm dis}},
 \end{align} 
where $t_0$ is a free parameter. The best-fit value obtained from our $N$-body models is $t_0/\Delta t\approx 0.028$, plotted in Fig.~\ref{fig:rate_tdis} with a black-dashed line for ease of reference.
Inserting Equation~(\ref{eq:ratefmax}) into~(\ref{eq:linear}) returns a dimension-less quantity
\begin{align}\label{eq:rate0}
  \xi\equiv t_{\rm dis}\,\mathcal{R}_{f,0}= 320\sqrt{\frac{15}{\pi}}\bigg(\frac{a}{r_c}\bigg)^{3/2}\bigg(\frac{m_\star }{M_c }\bigg)^{1/2}\frac{t_0}{\Delta t},
\end{align}
As we will see below, this number is directly proportional to the fraction of cluster particles that become bound pairs.

\begin{figure}
\begin{center}
\includegraphics[width=84mm]{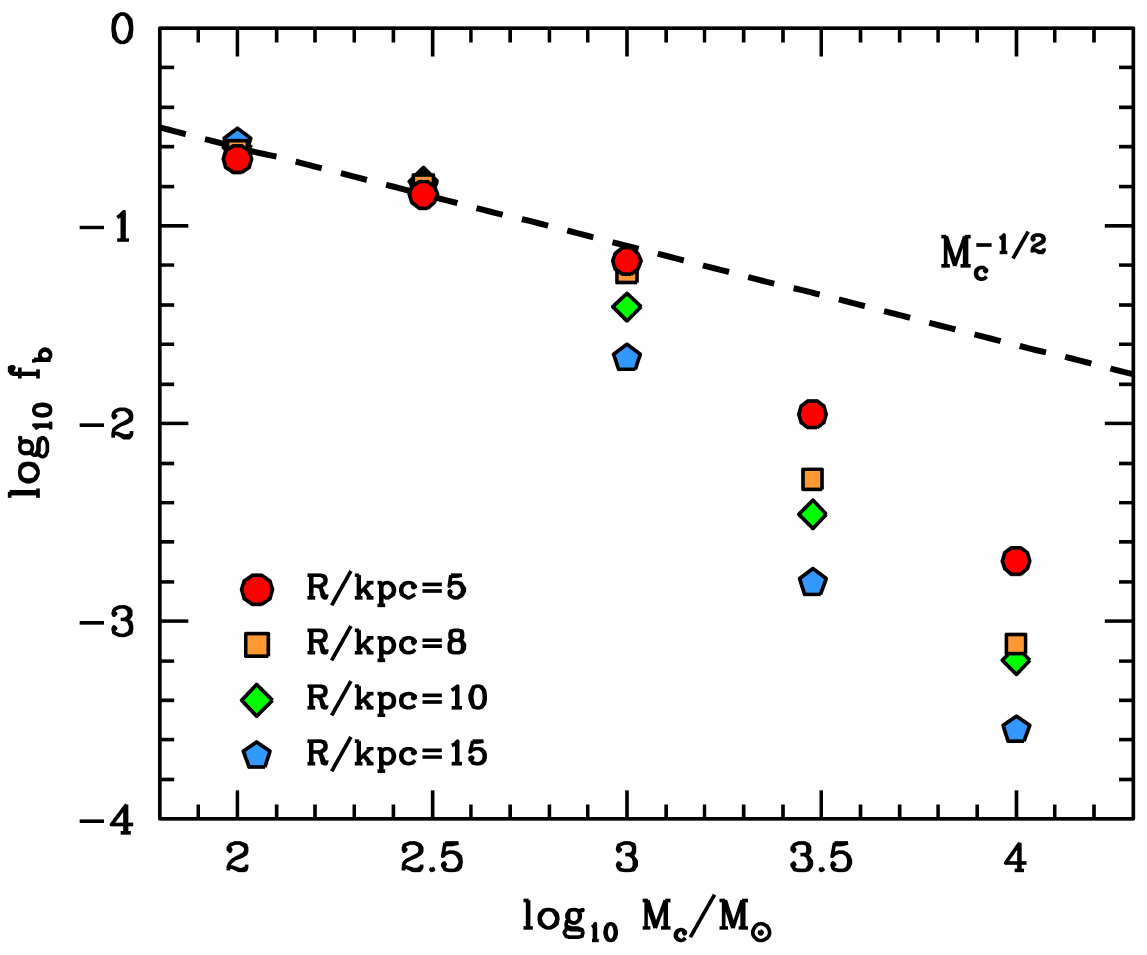}
\end{center}
\caption{Binary fraction, Equation~(\ref{eq:Nb}), as a function of cluster mass. Coloured symbols denote models with different orbital radii ($R$). Cluster $N$-body realizations follow Dehnen (1993) spheres with an initial scale-length $r_c=5\pc$ and are integrated $5\gyr$.  Note that clusters with large masses/orbital radii are not fully disrupted by the end of the simulation, which leads to a deficiency in the number of binaries with respect to the analytical prediction~(\ref{eq:fbfinal}) in the limit $t_{\rm dis}/t_{\rm now}\to 0$ (black-dashed line). As expected, models that are fully disrupted at early times follow a binary fraction that scales as $f_b\sim M_c^{-1/2}$ independently of their orbital radius. }
\label{fig:nbin}
\end{figure}

Indeed, time integrating the binary formation rate~(\ref{eq:Rf0t0}) and inserting~(\ref{eq:rate0}) yields
\begin{align}\label{eq:fbfinal}
  f_b&=\int_0^{t_{\rm now}}\d t\,\mathcal{R}_{f}(t)=\int_0^{t_{\rm dis}}\d t\,\mathcal{R}_{f}(t)+\int_{t_{\rm dis}}^{t_{\rm now}}\d t\,\mathcal{R}_{f}(t) \\ \nonumber
 &=\xi\bigg(2-\frac{t_{\rm dis}}{t_{\rm now}}\bigg)\\ \nonumber 
      & \approx 19.6\bigg(\frac{a}{r_c}\bigg)^{3/2}\bigg(\frac{m_\star }{M_c }\bigg)^{1/2}\bigg(2-\frac{t_{\rm dis}}{t_{\rm now}}\bigg),
\end{align}
for $t_{\rm dis}\le t_{\rm now}$ and $a\ll r_c$. Comparison of Equations~(\ref{eq:Q0}) and~(\ref{eq:fbfinal}) shows that the binary fraction is proportional to the maximum phase-space density of the cluster, i.e. $f_b\sim Q_0$.  
Furthermore, Equation~(\ref{eq:fbfinal}) implies that the number of binaries created {\it prior} to the disruption a stellar cluster is equal to that that will form {\it after} the cluster has been fully disrupted. This can be shown by integration of the formation rate~(\ref{eq:Rf0t0}) within the two relevant time intervals, which yields
$$\int_0^{t_{\rm dis}}\d t\,\mathcal{R}_{f}(t)=\int_{t_{\rm dis}}^\infty\d t\,\mathcal{R}_{f}(t)=\,t_{\rm dis}\,\mathcal{R}_{f,0}=\xi,$$
hence the total binary fraction scales as $f_b= 2\,\xi\sim M_c^{-1/2}$ (black-dashed line in Fig.~\ref{fig:nbin}).
As expected, the scale-free relation works well for clusters with low mass \& small orbital radii, which tend to have short disruption times, $t_{\rm dis}\ll t_{\rm now}$. In contrast, clusters with high mass \& large orbital radii are not fully disrupted within the integration time of our $N$-body experiments. As a result, the fraction of binaries formed in these models falls below the analytical expectation $f_b= 2\,\xi$ derived in the limit $t_{\rm dis}/t_{\rm now}\to 0$. 
Note also that Equation~(\ref{eq:fbfinal}) returns unphysical values $f_b>1$ for $M_c\lesssim 12\msol$, which suggests that the best-fit parameter $t_0/\Delta t$ should not be extrapolated to arbitrarily-low cluster masses.



The total number of bound stellar pairs associated with the disruption of a single cluster can be simply estimated from~(\ref{eq:fbfinal}) by setting $N_{\rm unb}=N_c=M_c/m_\star$ in Equation~(\ref{eq:Nb}) and inserting~(\ref{eq:fbfinal}), which yields
\begin{align}\label{eq:Nbfinal}
  N_b\approx 19.6\bigg(\frac{M_c }{m_\star }\bigg)^{1/2}\bigg(\frac{a}{r_c}\bigg)^{3/2}\bigg(2-\frac{t_{\rm dis}}{t_{\rm now}}\bigg).
\end{align}
Thus, on long time-scales $t_{\rm now}\gg t_{\rm dis}$, the number of wide binaries scales with the initial mean density of the progenitor cluster as $N_b\propto (M_c/r_c^3)^{1/2}\sim \rho_c^{1/2}$, independently of the cluster's orbital radius.


\begin{figure*}
\begin{center}
\includegraphics[width=160mm]{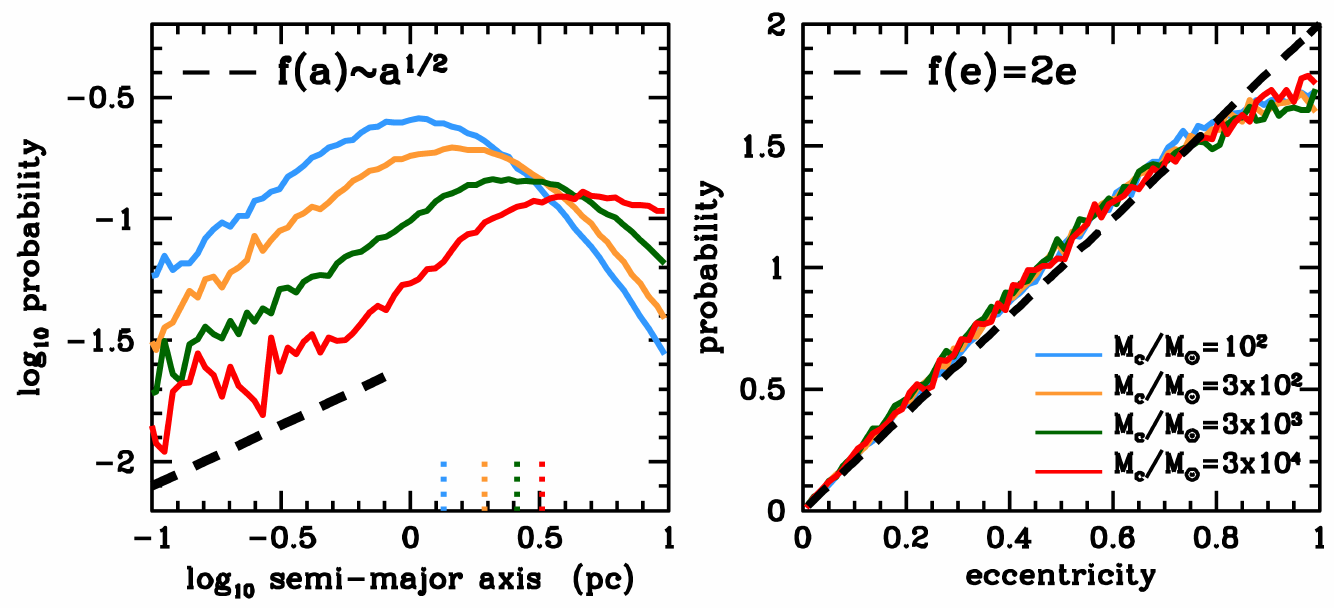}
\end{center}
\caption{{\it Left panel}: semimajor axis distribution of wide binaries formed in the tidal streams associated with tidally-disrupting clusters with different masses ($M_c$) and a fixed initial scale-length $r_c=5\pc$. The semimajor axis distribution peaks at a distance that approximately coincides with the average separation of stream stars, $a_{\rm peak}\sim D$ (marked with vertical dotted lines for reference). As expected, the theoretical curve~(\ref{eq:fa}), $p(a)\sim a^{1/2}$, fits well on small scales $a\ll D$ (shown with black-dashed line for reference). {\it Right panel}: Eccentricity distribution of wide binaries. Notice the scarcity of wide binaries on eccentric orbits with respect the thermal distribution $p(e)\d e =2e\d e$ (black-dashed line). } 
\label{fig:bin_2}
\end{figure*}

\subsection{Orbital elements of wide binaries }\label{sec:binary}
In isolation, bound pairs move in elliptical orbits around a Keplerian potential $\Phi_b=-Gm_b/r$.
The peri- and apocentres of the orbit correspond to the radii where the radial velocity cancels, $v_r=\{2[E-\Phi_b(r)-L^2/(2\,r^2)]\}^{1/2}=\{2[-G m_b/(2\,a)+G m_b/r-G m_b\,a(1-e^2)/(2\,r^2)]\}^{1/2}=0$, which admits two solutions: a pericentric radius $r_p=a(1-e)$, and apocentre $r_a=a(1+e)$, where $a=Gm_b/(-2E)$ is the semimajor axis and $e=\sqrt{1-L^2/(G m_b\, a)}$ is the eccentricity of the orbit. Here, $E=v^2/2-Gm_b/r<0$ is the specific energy and $\bb L=\bb r\times \bb v$ is the specific angular momentum of particle pairs with a combined mass $m_b=1\msol$ separated by a distance $\bb r$ and moving with a relative velocity $\bb v$ at the time of detection. To derive the average orbital radius one can use Appendix C of Paper II, which yields $\overline {r}=2/P\int_{r_p}^{r_a}\d r\,(r/v_r)=a(1+e^2/2)$ for an orbital period $P=2\int_{r_p}^{r_a}\d r/v_r=2\pi a^{3/2}/(Gm_b)^{1/2}$.

Fig.~\ref{fig:bin_2} shows the semimajor axis (left panel), and the eccentricity (right panel) of bound pairs found in the tidal debris of disrupting clusters with an initial scale length $r_c=5\pc$ and a mass $M_c$ moving on circular orbits at a galactocentric radius $R=5\pc$. All curves are normalized such that $\int_0^\infty \d a\,p(a)=\int_0^1 \d e\,p(e)=1$.
This figure illustrates a few interesting points. One is that the probability function $p(a)$ peaks at a semimajor axis $a_{\rm peak}$, such that $\d p/\d a|_{ a_{\rm peak}}=0$, which therefore defines the most-likely semimajor axis of the wide binary population. Note that the length of $a_{\rm peak}$ shifts to larger values as the mass of the cluster models increases. 
For cluster masses $\log_{10} M_c=2, 2.5, 3.5$ and 4.5 we find maxima at $a_{\rm peak}/\pc\simeq 1.0, 1.3, 2.1$ and $4.1$,
respectively. Hence, the location of the peak roughly scales as $a_{\rm peak}\sim  M_c^{1/3}$, the same power-law behaviour exhibited by the tidal radius of clusters on circular orbits, $r_t\sim M_c^{1/3}$ (e.g. Renaud et al. 2011).
These results are consistent with Fig.~\ref{fig:phase}, which shows that the majority bound pairs in $N$-body models have a separation that is comparable to the average distance of stream particles, $\overline {r}\sim D$, and that the number of these objects decreases sharply at larger ($r\gtrsim D$) and smaller ($r\lesssim D$) separations. Hence, it follows that wide binaries form in tidal streams with a semimajor distribution that peaks at the inter-stellar distance, $a_{\rm peak}\sim\overline {r}\sim D$.
A second point of interest is that the distribution converges to a scale-free function $p(a)\sim a^{1/2}$ at $a\ll a_{\rm peak}$ (marked with a black-dashed line), confirming the statistical expectation given by Equation~(\ref{eq:fa}). The power-law behaviour can be better seen in massive cluster models, which produce broad streams with large inter-stellar distances. 

Models shown in the right panel of Fig.~\ref{fig:bin_2} indicate that, independently of cluster mass, wide binaries form in tidal streams with an eccentricity distribution that is close to {\it thermal}, $p(e)\,\d e=2\,e\d e$ (black-dashed line), although a noticeable scarcity of wide binaries on eccentric orbits is visible at $e\gtrsim 0.8$. 
A thermalized eccentricity spectrum arises when orbiral energies follow a Boltzmann distribution (Jeans 1928; Heggie 1975)
\begin{align}\label{eq:boltz}
  f(E)\d E \sim \exp(-E/T)\frac{\d E}{(-E)^{5/2}},
  \end{align}
where $T$ is the mean kinetic energy. Note that for weakly-bound objects, $|E|\ll T$, Equation~(\ref{eq:boltz}) reduces to the energy distribution of random pairs with Maxwellian velocities, Equation~(\ref{eq:fE}), which is derived under the assumption that the a semimajor axis distribution scales as $p(a)\sim a^{1/2}$ at $a\lesssim D$.


\subsection{Age spread }\label{sec:age}
When do wide binaries form? According to Fig.~\ref{fig:rate_evol}, most pairs become gravitationally bound at early times, when mass-loss rates are high and stream particles are still in the vicinity of the cluster (see Fig.~\ref{fig:Q}).
At this early stage, the {\it entrapment} of random pairs predominantly occurs as particles escape through the Lagrange points and the progenitor's tidal field acting on them weakens, progressively vanishing as particles drift away along the tidal tails. 
This simple picture changes dramatically once the cluster has been fully disrupted. At this later stage, the mechanism that dominates 
the coalescence of bound pairs is the random conjunction of stream particles that may have been stripped at very different times. Given that massive clusters tend to exhibit gradients in age \& composition (e.g. Bastian \& Lardo 2018), this process may lead to the formation of wide binaries with a measurable spread of elements\footnote{Interestingly, comoving pairs with dissimilar chemical composition are not uncommon (e.g. Ram\'irez et al. 2019).}.

 
 To inspect this issue in more detail, we record the time $t_{\rm unb}$ at which $N$-body particles become energetically-unbound from the progenitor cluster and compute the 'age difference' between the primary and companion star, $\Delta t_{\rm unb}=t_{\rm unb,2}-t_{\rm unb,1}$. The unbinding time is defined within the interval $t_m\le t_{\rm unb}\le t_{\rm dis}$, where $t_{\rm m}> 0$ is the time at which a cluster begins to shed mass to tides. 
 Clearly, the distribution of unbinding times must be related to the rate at which particles are tidally stripped from the progenitor cluster. Following the results of Fig.~\ref{fig:rate_evol}, let us model the fractional mass-loss rate with a truncated power-law function
 \begin{eqnarray}\label{eq:Runb}
  \mathcal{R}_{\rm unb}(t)=
  \begin{cases}
    \mathcal{R}_{0}\big(t_{\rm m}/t)^{\alpha} &~~\,t_m\le t\le t_{\rm dis},\\
     0 &~~{\rm otherwise.}
    \end{cases}
 \end{eqnarray}
 with $\alpha\approx 1.2$. 
 The normalization factor $\mathcal{R}_{0}$ follows from integration of Equation~(\ref{eq:Runb}) over the life-span of the cluster
 \begin{align}\label{eq:R0}
   f_{\rm unb}(t_{\rm dis})=\int_{t_m}^{t_{\rm dis}}\d t'\,\mathcal{R}_{\rm unb}(t')= \frac{\mathcal{R}_0 \,t_m}{\alpha-1}\bigg[1-\bigg(\frac{t_m}{t_{\rm dis}}\bigg)^{\alpha-1}\bigg]=1,
 \end{align}
 In general, it is safe to assume that clusters do not dissolve immediately in the host tidal field. Hence, we can approximate $t_m\ll t_{\rm dis}$ in Equation~(\ref{eq:R0}), which leads to a normalization factor $\mathcal{R}_0\approx (\alpha-1)\,t_m^{-1}$. Using this normalization allow us to treat the fractional mass-loss rate $\mathcal{R}_{\rm unb}(t)$ as the probability density to find stream particles stripped in the time-interval $t,t+\Delta t$. 

 Let us now turn to the problem of computing the probability that {\it two} random stars form a bound pair with an age spread $\Delta t_{\rm unb}=t_{\rm unb,2}-t_{\rm unb,1}$. We start by simplifying our notation and introducing the variable $\tau\equiv t_2-t_1$, where $t_i=t_{\rm unb,i}$ and $i=1,2$. The quantity of interest is the joint probability $p(t_2=\tau)\cdot p(t_2=t_1+\tau)$, which corresponds to the {\it autocorrelation} of the fractional mass-loss rates given by Equation~(\ref{eq:Runb}), that is $p(\tau)=\int_{-\infty}^{+\infty}\d t\,\mathcal{R}_{\rm unb}(t)\mathcal{R}_{\rm unb}(\tau+t)$. Particles 1 and 2 can be exchanged without loss of generality, which means that the autocorrelation function must be even, i.e. $p(\tau)=p(-\tau)$. For simplicity, let us consider the case $\tau>0$. The limits of the integral are set by the range of values of $t_1$ and $t_2$, namely $t_m\le t_1\le t_{\rm dis}$ and $t_2=t_1+\tau$. 
 Under the condition $t_m\ll t_{\rm dis}$, the autocorrelation function can be expressed in an analytical form  by taking the upper limit of the integral to $t_{\rm dis}\to \infty$ and Taylor expanding the result in series of $t_m$, which yields
\begin{align}\label{eq:pdeltat}
  p(\tau)&=\int_{t_m}^{t_{\rm dis}}\d t\,\mathcal{R}_{\rm unb}(t)\mathcal{R}_{\rm unb}(t+\tau) \\ \nonumber
  &\approx \mathcal{R}_0^2 \,t_m^{2\alpha}\int_{t_m}^{\infty}\frac{\d t}{t^{\alpha}(\tau+t)^{\alpha}}\\ \nonumber
  &=\mathcal{R}_0^2 \,t_m^{2\alpha}\bigg[\frac{1}{t_m^{\alpha-1}\tau^\alpha}\frac{\Gamma(\alpha-1)}{\Gamma(\alpha)}+\frac{1}{\tau^{2\alpha-1}}\frac{\Gamma(1-\alpha)\Gamma(2\alpha)}{(2\alpha-1)\Gamma(\alpha)}+\mathcal{O}(t_m^2)\bigg],
\end{align}
where $\Gamma(x)$ is the Gamma function. 
Inserting the normalization factor $\mathcal{R}_0= (\alpha-1)\,t_m^{-1}$ and taking the leading order in the limit $t_m\to 0$ yields
\begin{align}\label{eq:pdeltat2}
   p(|\tau|)\simeq \frac{(\alpha-1)^2}{2}\frac{\Gamma(\alpha-1)}{\Gamma(\alpha)} \frac{t_m^{\alpha-1}}{|\tau|^{\alpha}}~~~~{\rm for}~~~|\tau|\ll t_{\rm dis},
\end{align}
where we have multiplied~(\ref{eq:pdeltat}) by $1/2$ to account for the area in the interval $\tau<0$, such that $\int_{-\infty}^{+\infty}\d\tau\,p(|\tau|)=2\int_{t_m}^{+\infty}\d\tau\,p(\tau)=1$.
 Equation~(\ref{eq:pdeltat2}) reveals a few important points. The first one is that the age spread function follows a power-law distribution that has the same index as the mass loss rate~(\ref{eq:Runb}), i.e. $p(|\Delta t_{\rm unb}|)\sim |\Delta t_{\rm unb}|^{-\alpha}$. This implies that clusters with steep mass-loss rates form wide binaries with a narrow spread of unbinding times, and vice versa. 
The second is that the age spread peaks at small ages $|\Delta t_{\rm unb}|\approx t_m$. The fact that the cluster models described in \S\ref{sec:setup} are tidally-filled means that mass stripping begins at early times, $t_m\approx 0$. As a result, the majority of bound pairs in these models are expected to exhibit small age spreads.



\begin{figure}
\begin{center}
\includegraphics[width=84mm]{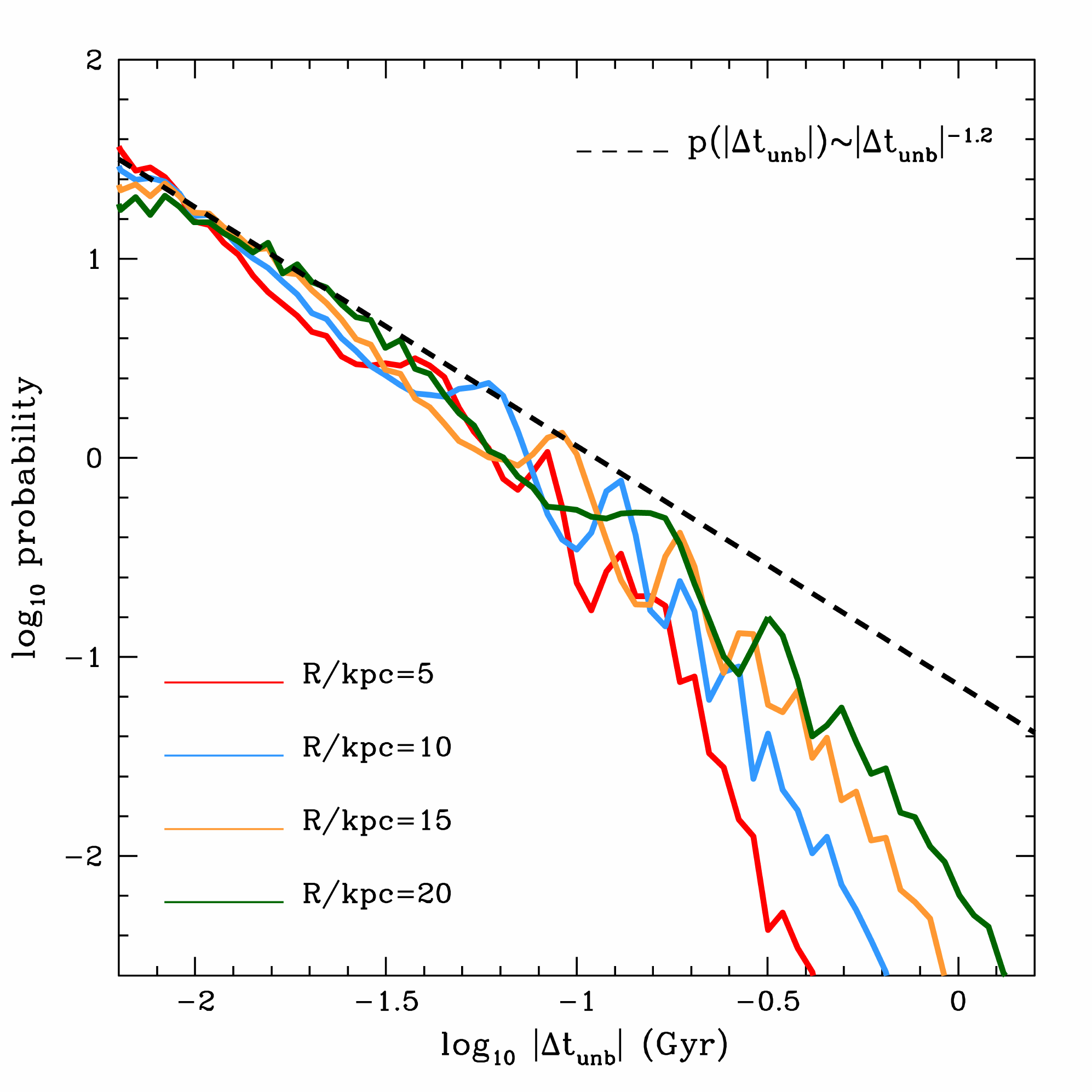}
\end{center}
\caption{`Age spread' of bound pairs, $|\Delta t_{\rm unb}|$, found in the $N$-body models of Fig.~\ref{fig:rate_evol}. Here, $t_{\rm unb}$ corresponds to the time at which a particle is tidally stripped from the progenitor cluster. The theoretical expectation~(\ref{eq:pdeltat2}) with $\alpha=1.2$ and $t_m=0.2\gyr$ is plotted with a black-dashed line. }
\label{fig:tm}
\end{figure}

Fig.~\ref{fig:tm} shows that these prediction largely agrees with the distributions of unbinding times measured in our $N$-body models. Inserting the power-law index $\alpha=1.2$ found in Fig.~\ref{fig:rate_evol} into Equation~(\ref{eq:pdeltat2}) yields $p(|\Delta t_{\rm unb}|)=0.1\,t_m^{0.2}\,|\Delta t_{\rm unb}|^{-1.2}$. Setting $t_m\approx 0.2\gyr$ in~(\ref{eq:pdeltat2}) provides a reasonable description of the $N$-body curves on time-scales much shorter than the disruption time of the progenitor, $|\Delta t_{\rm unb}|\ll t_{\rm dis}$. Recall that clusters orbiting in the outskirts of the Galaxy potential have longer disruption times (see Fig.~\ref{fig:rate_evol}). This leads to a power-law behaviour $p(|\Delta t_{\rm unb}|)\sim |\Delta t_{\rm unb}|^{-1.2}$ that extends over larger age intervals as the orbital radius $R$ increases.


\section{Survival of wide binaries in a clumpy galaxy potential}\label{sec:disrup}
Thus far our models have ignored the effects of the external tidal field on the properties of wide binaries. As a result, the statistical model presented in Section~\ref{sec:model} makes unrealistic predictions on the number and distribution of wide binaries at separations which --in theory-- can reach arbitrarily large values. In practice, such loosely-bound objects survive for a very short time in the galactic tidal field (Heggie 1975). To inspect this issue in more detail, this Section analyzes the dynamical evolution of bound pairs found in $N$-body simulations of tidally-disrupting clusters which are acted on by an external tidal field that contains two components: a term that changes very slowly and can be expressed as the gradient of a smooth potential, plus a random contribution of ``chance stellar encounters'' of short duration.

In the {\it tidal approximation}, the relative motion between a stellar pair can be described with the following equations of motion (see P19a for details)
\begin{align}\label{eq:eqmots}
\frac{{\d^2 \mathbfit r}}{\d t^2}=-\frac{Gm_b}{r^3}{\mathbfit r} + T_g\cdot {\mathbfit r} + \sum_{i=1}^{N} t_i\cdot {\mathbfit r},
\end{align}
here, ${\mathbfit r}={\mathbfit r}_A-{\mathbfit r}_B$ is the relative separation between two mutually-gravitating particles A and B that have been tidally-stripped from a progenitor cluster, and $T_g$ and ${t}_i$ are 3$\times$3 tidal tensors evaluated at the barycentre of the pair. The smooth component has a form
\begin{align}\label{eq:tt_s}
  {T}_{g}^{jk}\equiv -\frac{\partial^2 \Phi_g}{\partial x_j\partial x_k}, 
\end{align}
where $\Phi_g$ is the mean-field gravitational potential of the host, while the stochastic tidal tensor  
\begin{align}\label{eq:tt}
 {T}^{jk}\equiv  \sum_{i=1}^{N}{t}_i^{jk}= \sum_{i=1}^{N} \frac{\partial f^k_{i}}{\partial x^j}=\frac{\partial }{\partial x^j}\sum_{i=1}^{N} f^k_{i}=\frac{\partial F^k}{\partial x^j},
\end{align}
arises from the gradient in the combined tidal force generated by a set of $N-${substructures} distributed across the host galaxy, which induce a specific force 
${\mathbfit F}\equiv \sum_{i=1}^{N}{\mathbfit f}_i$. 
Following P19a, we will assume that substructures follow Hernquist (1990) density profiles with a mass $M$ and a scale-radius $c$,  which generate individual forces
\begin{align}\label{eq:fhern}
  {\mathbfit f}_i=-\frac{G M}{(R'_i+c)^2}\hat R'_i,
\end{align}
where ${\mathbfit R}'_i={\mathbfit R}_b-{\mathbfit R}_i$ is the relative distance between the binary barycentre and the $i^{th}$ substructure. 
In this paper we only consider the contribution of disc stars to the random component of the tidal field, thus neglecting the presence of dark substructures and molecular clouds in the galaxy. We will come back to this point in \S\ref{sec:discuss}.

\subsection{Smooth tidal field}\label{sec:smooth}
Before integrating the differential equations~(\ref{eq:eqmots}), it is useful to estimate the separation at which the mutual attraction
between a bound pair becomes comparable to the strength of the smooth
tidal field. This can be done by computing the tidal radius of a binary star moving on circular orbits in the disc plane as (see Pe\~narrubia et al. 2016 for details)
\begin{eqnarray}\label{eq:rt}
r_t(R)\equiv \bigg[\frac{G m_b}{\lambda_1}\bigg]^{1/3},
\end{eqnarray}
where 
\begin{eqnarray}\label{eq:lambda}
\lambda_1(R)=-\frac{\partial^2\Phi_g}{\partial x^2}\bigg|_{{\bf R}}+\frac{\partial^2\Phi_g}{\partial z^2}\bigg|_{{\bf R}}\approx \gamma\Omega_g^2,
\end{eqnarray}
is the eigenvalue of the effective tidal tensor, $T_e^{ij}=-\partial^2(\Phi_g+\Phi_c)/(\partial x_i\partial x_j)$, which contains a centrifugal potential $\Phi_c=-\Omega_g^2r^2/2$. Here, $\Omega_g(R)=v_c(R)/R=[G M_g(<R)/R^3]^{1/2}$ is the circular frequency of the pair about a host galaxy with a mass profile $M_g(R)$, and $\gamma(R)\equiv -\d \log \rho/\d \log r$ is the power-law slope
of the host's density profile computed at the galactocentric radius $R$ (Renaud
et al. 2011). Note that in a Keplerian potential $\gamma=3$ and
$\Omega_g^2=GM_g/R^3$, which recovers the well-known Jacobi radius 
$r_t=R[m_b/(3M_g)]^{1/3}$. For the extended Milky Way-like models introduced in \S\ref{sec:setup}, the density slope ranges from $\gamma(1\kpc)\simeq 0.49$ up to $\gamma(20\kpc)\simeq 2.92$ in the outskirts of the disc.

\subsection{Clumpy tidal field}\label{sec:clumpy}
In addition to the mean-field potential of the host galaxy, the survival of wide binaries also depends on the `granularity' of the local mass distribution.
Stellar pairs moving in a clumpy medium experience stochastic fluctuations of the tidal field due to the rapid change of the (relative) position of nearby substructures.
Chandrasekhar (1941a,b; 1943) argues that the cumulative effect of force fluctuations leads to random increments of the particle velocity, $\Delta {\mathbfit v}$, which can be treated as a random walk in a three-dimensional velocity space.
In this theoretical framework, the distribution of velocity impulses that is isotropic and has a Gaussian form (Chandrasekhar 1943; Kandrup 1980)
\begin{align}\label{eq:Psi}
 \Psi({\mathbfit v}, \Delta {\mathbfit v},t-t_0)=\frac{1}{(\frac{2\pi}{3}\langle |\Delta{\mathbfit v}|^2\rangle)^{3/2}}\exp\big[-\frac{(\Delta{\mathbfit v}-\langle \Delta{\mathbfit v}\rangle)^2}{\frac{2}{3} \langle |\Delta{\mathbfit v}|^2\rangle}\big];
\end{align}
where $\Psi({\mathbfit v}, \Delta {\mathbfit v},t-t_0)$ denotes the probability that a test particle with a velocity ${\mathbfit v}$ will experience a velocity impulse $\Delta{\mathbfit v}$ within a time interval $t-t_0$. Unless otherwise indicated, we set $t_0=0$ for simplicity.

The probability function~(\ref{eq:Psi}) is uniquely defined by two coefficients, $\langle \Delta {\mathbfit v}\rangle$ and $\langle |\Delta {\mathbfit v}|^2\rangle$, with brackets denoting averages over multiple fluctuations. If one assumes that nearby substructures are isotropically distributed around the stellar pair, then the first moment $\langle \Delta {\mathbfit v}\rangle=0$ by symmetry. The second moment is more difficult to compute, and generally depends on the ratio between the characteristic duration of a tidal fluctuation, $T_{\rm ch}=0.88\,D/\sqrt{\langle v^2\rangle}$, and the orbital frequency of the bound pair, $w=v/r$. Here, $\sqrt{\langle v^2\rangle}$ is the relative speed of substructures separated by a mean distance $D$.
In the {\it local approximation}, where the number density of substructures, $n=(2\pi D^3)^{-1}$, is assumed to be constant within a distance scale $r\lesssim d\equiv |\nabla n/n|^{-1}$, the diffusion coefficient can be written as (see P19a for details)
\begin{align}\label{eq:dv2}
  \langle |\Delta {\mathbfit v}|^2\rangle=\delta_s(t)\times \langle |\Delta {\mathbfit v}|^2\rangle_\infty,
\end{align}
where
\begin{align}\label{eq:dv2inf}
  \langle |\Delta {\mathbfit v}|^2\rangle_\infty=
  \begin{cases}
 t\,r^2 \frac{4\pi}{5}\frac{(GM)^2}{c^2}n{\sqrt\frac{2\pi}{3 \langle v^2\rangle}}& , T_{\rm ch}\ll w^{-1} ~~({\rm impulsive}) \\ 
 t\,\frac{r^5}{v^3}\frac{24\pi}{5} \frac{(GM)^2}{c^5}n\langle v^2\rangle   & , T_{\rm ch}\gg w^{-1} ~~({\rm adiabatic})    
\end{cases}
\end{align}
is the analytical expression obtained in a Brownian motion framework and
\begin{align}\label{eq:deltas}
  \delta_s(t)=1-\exp(-t\, w_s),
 \end{align} 
is the so-called {\it sampling delay} function, which is determined by the {\it sampling frequency}
\begin{align}\label{eq:omega_s}
w_s\approx 1.49 \,c^2 n\sqrt{\langle v^2\rangle}.
\end{align}
The function~(\ref{eq:deltas}) is an empirical correction due to the fact that, although very close encounters with compact ($c\ll D$) substructures are very rare, on average they provide the largest contribution to the velocity impulses.
Direct-force $N$-body experiments carried by P19a show that the analytical expression~(\ref{eq:dv2}) corresponds to the asymptotic behaviour of $\langle |\Delta {\mathbfit v}|^2\rangle$ on long time-scales, $t\,w_s\to \infty$ ($\delta_s\to 1$). If the time-interval is short, $t\,w_s\lesssim 1$, the probability to sample strong-force events is low, which results in an ensemble-average amplitude of $\langle |\Delta {\mathbfit v}|^2\rangle$ that is systematically suppressed with respect to the random-walk value (see Fig.3 of P19a).

The coefficient derived from random-walk statistics, $\langle |\Delta {\mathbfit v}|^2\rangle_\infty$, exhibits two different behaviours depending on the average duration of the tidal fluctuations relative to the orbital period of the binary. At a fixed orbital frequency, $w=v/r$, the divide between the impulsive and adiabatic regimes is largely set by the distance-to-size ratio of substructures. This can be seen by equating the adiabatic and impulsive terms in Equation~(\ref{eq:dv2}), which returns a transition frequency
\begin{align}\label{eq:wad}
 w_{\rm ad}\approx 1.41\bigg(\frac{D}{c}\bigg)T_{\rm ch}^{-1}.
\end{align}
Point-mass particles ($c/D\to 0$) have a divergent transition frequency, $w_{\rm ad}\to \infty$, which indicates that the velocity impulses generated by compact objects must be treated impulsively. In this work we are mainly interested in tidal fluctuations induced by compact objects, such as stars in the solar neighbourhood, which have individual sizes $c\sim 10^{-8}\pc$ and an average separation $D\sim 1\pc$. Hence, the remainder of this paper works under the impulse approximation.

Very compact objects ($c/D\ll 1$) have a sampling frequency~(\ref{eq:omega_s}) that vanishes as $w_s\sim (c/D)^2 T_{\rm ch}^{-1}\to 0$ in the limit $c/D\to 0$. This calls for Taylor-expanding the sampling delay function~(\ref{eq:deltas}) at leading order, $\delta_s(t)\approx t\,w_s$. Inserting~(\ref{eq:omega_s}) into the impulsive term of Equation~(\ref{eq:dv2}) returns a coefficient
\begin{align}\label{eq:dv2_imp}
  \langle |\Delta {\mathbfit v}|^2\rangle&=
  t^2\,\,r^2 \frac{4\pi}{5}\frac{(GM)^2}{c^2}n{\sqrt\frac{2\pi}{3 \langle v^2\rangle}}\,w_s\\ \nonumber
  &\approx 5.42 \,t^2 r^2 (G M)^2n^2 \\ \nonumber
  &=5.42 \,t^2 r^2 (G\rho)^2 ~~~~~{\rm for}~~~c/D\ll 1,
\end{align}
where $\rho=M\,n$ is the mean density of substructures. It is remarkable that the diffusion coefficient associated with compact objects is independent of their relative velocity ($\langle v^2\rangle$), and their individual masses ($M$) and sizes ($c$) so long as their mean density is fixed.


In an impulse regime, the relative location of binary stars is assumed to remain constant during a tidal fluctuation. Hence, the variation of orbital energy is equal to the change of kinetic energy
\begin{align}\label{eq:delE}
   \Delta E=\frac{1}{2}({\mathbfit v}+\Delta {\mathbfit v})^2-\frac{1}{2}{\mathbfit v}^2={\mathbfit v}\cdot\Delta{\mathbfit v}+\frac{1}{2}(\Delta {\mathbfit v})^2.
\end{align}
If one assumes that the population of substructures are isotropically distributed around the binary, then the average over multiple fluctuations of the first right-hand term is $\langle {\mathbfit v}\cdot\Delta{\mathbfit v}\rangle =0$ by symmetry. Inserting~(\ref{eq:dv2_imp}) in~(\ref{eq:delE}) and taking the average over binary ensembles with fixed orbital energy and angular momentum yields 
\begin{align}\label{eq:delE2}
  \langle \Delta E\rangle (a,e)&=\frac{1}{2}\langle|\Delta {\mathbfit v}|^2\rangle \\ \nonumber &=2.71\,t^2(G\rho)^2 \overline{r^2}\\ \nonumber
  &=2.71\,t^2(G\rho)^2 a^2\bigg(1+\frac{3}{2}e^2\bigg),
\end{align}
where the last equality uses the ensemble-averaged distance squared of binary stars with a fixed semimajor axis and orbital eccentricity, $\overline{r^2}=a^2(1+3e^2/2)$ (see Appendix C of P19a). 

The positive sign of Equation~(\ref{eq:delE2}), $\langle \Delta E\rangle >0$, implies that on average binary systems absorb kinetic energy from the fluctuating tidal field. Over time, this process causes a progressive unbinding of self-gravitating objects, which is typically known as ``tidal evaporation'' (Spitzer 1958 and references therein).
The characteristic lifetime of a binary with orbital energy $E_0=-Gm_b/(2a)<0$ at $t=0$ can be grossly estimated from Equation~(\ref{eq:delE2}) by equating $\langle \Delta E\rangle=0-E_0$, which yields\footnote{For simplicity, we have neglected the contribution of the diffusion term $\langle \Delta E^2\rangle$ to Equation~(\ref{eq:tesc}). A more accurate estimate of the escape time can be obtained by computing the average time that it takes for particles with an energy distribution $\delta(E-E_0)$ at $t=0$ to reach the boundary $E=0$. E.g. see \S6.2 of P19a.} 
\begin{align}\label{eq:tesc}
  t_{\rm esc}(a,e)\approx 0.43\,\frac{m_b^{1/2}}{G^{1/2}\rho \,a^{3/2}}\frac{1}{(1+3e^2/2)^{1/2}}.
\end{align}
The difference between radial ($e=1$) and circular ($e=0$) orbits is a factor $t_{\rm esc}(e=1)/t_{\rm esc}(e=0)=(2/5)^{1/2}\approx 0.63$, which suggests that binary stars on eccentric orbits are more vulnerable to tidal fluctuations. Note also that dropping the eccentricity dependence $(1+3e^2/2)^{-1/2}$ from~(\ref{eq:tesc}) yields an expression similar to the dissolution time derived by Chandrasekhar (1944) from differential effects of force fluctuations acting on neighboring stars, the half-life time obtained by Bahcall et al. (1985) in the catastrophic regime where single collisions with pointlike objects break up the binary, and the characteristic life time obtained by Wielen (1985) and Weinberg et al. (1987) due to diffusive, non-penetrating encounters with field objects (which is the scenario adopted here).

In the Brownian motion theory, the average amplitude of energy impulses increases with the length of the time interval, $t$. This means that over sufficiently-long time-scales the cumulative effect of local tidal fluctuations is bound to dominate the destruction of wide binaries. To inspect this issue in more detail, it is useful to define a time-scale $t_{\rm sto}(R)\equiv \int_0^1\d e\,p(e)\, t_{\rm esc}(r_t,e)$, such that for long time intervals $t\gg t_{\rm sto}$ disruption is driven by tidal evaporation, whereas for $t\lesssim t_{\rm sto}$ the effect of clumpiness can be neglected. Here, $p(e)$ is the eccentricity distribution of wide binaries, which can approximated by a thermal distribution following the results plotted in Fig.~\ref{fig:bin_2}. Inserting the tidal radius~(\ref{eq:rt}) in Equation~(\ref{eq:tesc}), and averaging over eccentricity $\int_0^1\d e\,2e (1+3e^2/2)^{-1/2}=2(-2+\sqrt{10})/3\simeq 0.775$ returns a time-scale
\begin{align}\label{eq:tsto}
  t_{\rm sto}\simeq \frac{\gamma^{1/2}}{3}\frac{\Omega_g}{G\rho},
\end{align}
which is independent of the binary mass and inversely proportional to the local density of substructures. To gain physical insight on this result, let us consider the case of self-gravitating power-law galaxies made of compact objects which follow cored ($\gamma=0$) and cuspy ($\gamma=1$) density profiles. The density and mass distribution can be written as $\rho=\rho_0(r/r_0)^{-\gamma}$ and $M(<r)=M_0(r/r_0)^\Gamma$, respectively, with $M_0=4\pi\rho_0r_0^3/(3-\gamma)$ and an index $\Gamma=3-\gamma$. 
Inserting these expressions in~(\ref{eq:tsto}) and multiplying by the circular frequency $\Omega_g^2=GM(<r)/r^3$ returns a stochastic time
\begin{align}\label{eq:tsto_om}
  t_{\rm sto}\,\Omega_g=\frac{4\pi}{3}\frac{ \gamma^{1/2}}{(3-\gamma)},
\end{align}
which highlights a few interesting aspects of the problem: first, note that in a power-law galaxy the stochastic time is a fixed fraction of the dynamical time. For logarithmic density slopes $0<\gamma\lesssim 0.39$ the stochastic time-scale is shorter than the local dynamical time, $t_{\rm sto}<\Omega_g^{-1}$, whereas in cuspy galaxies with $0.39\lesssim \gamma<1$ the situation reverses. An interesting limiting case corresponds to galaxies with a cored profile, $\gamma=0$, for which the stochastic time-scale vanishes. This is a consequence of the compressive tidal field of homogeneous systems, which leads to a divergent tidal radius in the limit $\gamma\to 0$ (Renaud et al. 2011), suggesting that in galaxies with shallow density profiles tidal evaporation is the main mechanism driving the disruption of wide binaries (see also Pe\~narrubia et al. 2016).

Given the above results, a marked bifurcation in the evolution of binaries moving in smooth \& clumpy potentials is expected on time-scales $t\sim t_{\rm sto}$, an issue analyzed below with the aid of Monte-Carlo $N$-body experiments.

\begin{figure*}
\begin{center}
\includegraphics[width=160mm]{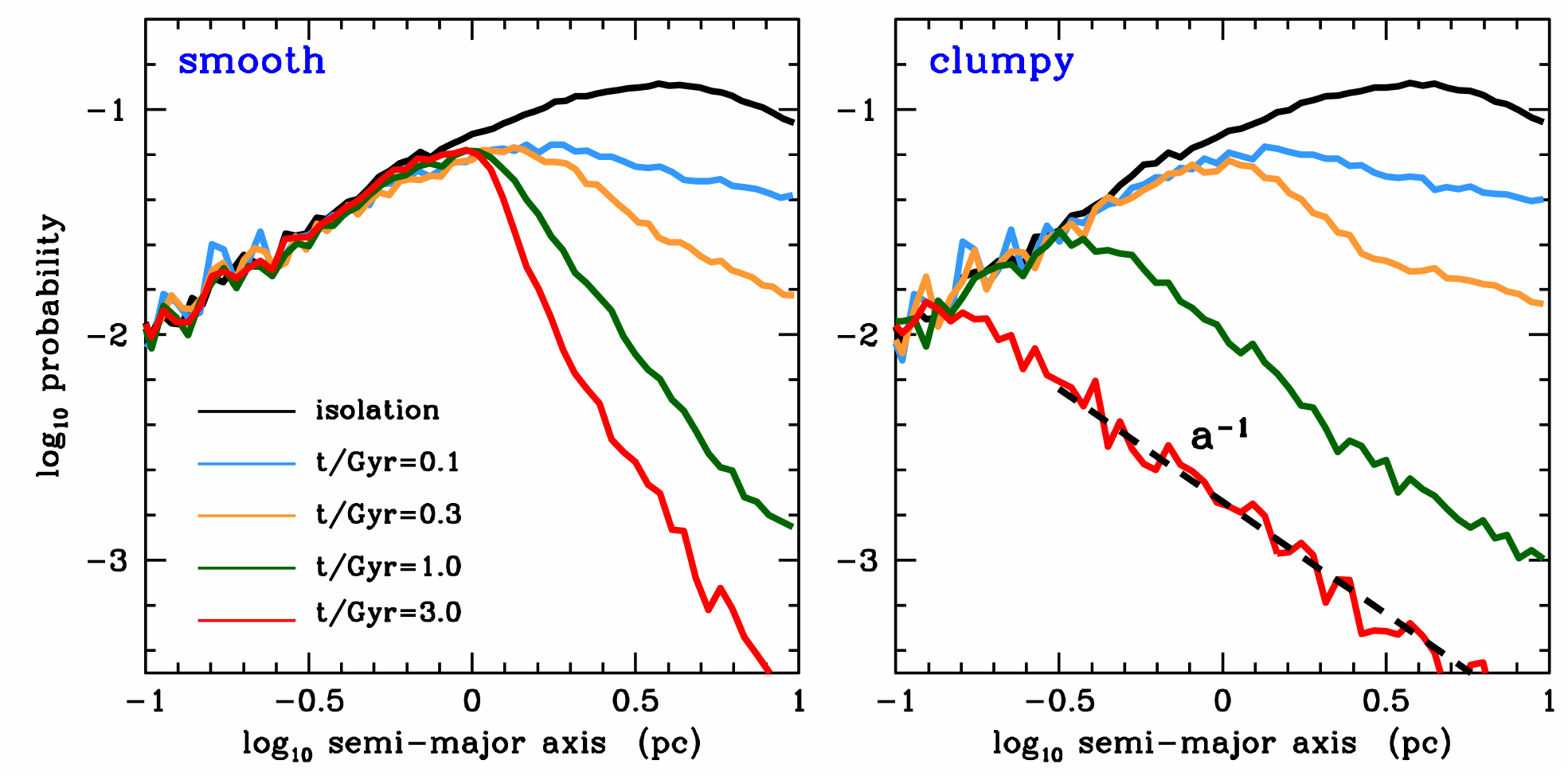}
\end{center}
\caption{{\it Left panel}: Distribution of semimajor axes of wide binaries with $m_b=1\msol$ orbiting in a ``smooth'' Milky Way potential at different time snapshots. Orbits around the host galaxy and the initial binary parameters are extracted from pairs found in the tidal debris of a cluster with an initial mass $M_c=300\msol$ and size $r_c=5\pc$ moving on a circular orbit at $(R,z)=(20,0)\kpc$ from the galactic centre. For reference, black lines show the distribution of bound pairs formed in isolation. The smooth tidal field wipes out binary stars with large semimajor axes (low-binding energies). As time proceeds, the distribution becomes sharply truncated at the tidal radius of the binary $a_{\rm peak}\approx r_t(20\kpc)=0.98\kpc$. {\it Right panel}: Distribution of semimajor axes of the same binaries plotted in the right panel, this time orbiting in a ``clumpy'' galactic model that mimics the tidal impulses generated by individual stars (see text). Note that on short times-scales, $t\lesssim t_{\rm sto}\approx 0.1\gyr$ the semimajor axis distributions in smooth and clumpy models are very similar. At later times, $t\gg t_{\rm sto}$, binaries moving in a clumpy medium develop a low-energy tail that scales as $p(a)\sim a^{-1}$, which is typically known as \"Opik (1924)'s law. As time goes by, the peak of the distribution becomes systematically smaller than the Jacobi radius, $a_{\rm peak}\ll r_t$.} 
\label{fig:a_2}
\end{figure*}

\subsection{Monte-Carlo $N$-body models}\label{sec:MC}
This Section studies the dynamical evolution of bound particle pairs identified in the numerical models of \S\ref{sec:nbody} orbiting in smooth and clumpy potentials. Our chief aim is to understand the role of compact substructures in driving the properties of wide binaries on long time-scales, $t\gg t_{\rm sto}$. In particular, we focus on the effect of random tidal forces caused by neighbouring stars.
Under the tidal approximation, the problem reduces to solving two sets of differential equations: (i) the equations of motion for the binary barycentre in a ``smooth'' galactic potential
\begin{align}\label{eq:mc}
\frac{{\d^2 \mathbfit R}}{\d t^2}=-\nabla \Phi_g.
\end{align}
and (ii) the reduced-mass equations for the relative separation between the bound stellar pair 
\begin{align}\label{eq:mc}
\frac{{\d^2 \mathbfit r}}{\d t^2}=-\frac{Gm_b}{r^3}{\mathbfit r} + T_g\cdot {\mathbfit r} ,
\end{align}
with $T_g=T_g({\mathbfit R})$ is the tidal tensor associated with the mean-field potential of the host galaxy. The initial conditions, $\{{\mathbfit R},\mathbfit V, {\mathbfit r},{\mathbfit v}\}(t_0)$, are extracted from particle pairs identified at different snapshots $t=t_0$ in the $N$-body models of \S\ref{sec:nbody}. Each individual pair is followed from its formation time $t_0$ until $t_{\rm now}=5\gyr$. Here, it is important to stress that the integration continues even when the binding energy turns positive, $E>0$, as formally disrupted binaries can be brought back to a bound configuration by the fluctuating tidal field (Jiang \& Tremaine 2010).

To model the effect of a ``clumpy'' host potential we inject uncorrelated velocity impulses at subsequent time-steps of the orbital integration\footnote{Interested readers are referred to P19a,b for an extensive study of this technique using controlled tests against direct-force experiments.}
 $${\mathbfit v}(t)\to {\mathbfit v}(t+\Delta t) + {\it Ran}(\Delta{\mathbfit v})$$
here, $\Delta t$ is the integration time-step, ${\mathbfit v}$ is the velocity vector computed from~(\ref{eq:mc}), and ${\it Ran}(\Delta{\mathbfit v})$ are random velocity increments drawn from the probability function $\Psi({\mathbfit v}, \Delta {\mathbfit v},\Delta t)$ given by~(\ref{eq:Psi}). Note that the length of the time interval between two consecutive time-steps in Equation~(\ref{eq:dv2inf}) is $\Delta t$, while the sampling delay function~(\ref{eq:deltas}) depends on the full integration time, $t$. Accordingly, the second moment of $\Psi$ associated with field stars can be computed from~(\ref{eq:dv2_imp}) as $\langle |\Delta {\mathbfit v}|^2\rangle =5.42 \,t\,\Delta t\, r^2 (G\rho_\star)^2$.

Fig.~\ref{fig:a_2} shows snapshots of 
the semimajor axis distribution of wide binaries evolving in ``smooth'' and ``clumpy'' galaxy potentials (left and right panels, respectively). As initial conditions we use bound pairs formed in the tidal tails of a cluster with an initial mass $M_c=300\msol$ and size $r_c=5\pc$ moving on a circular disc orbit at $R=20\kpc$. Fig.~\ref{fig:rate_evol} shows that (1) the cluster has not been fully disrupted by $t_{\rm now}=5\gyr$ (upper panel), and (2) this model produces bound pairs at an approximately constant rate throughout the cluster dissolution (lower panel). At this galactocentric radius, the stochastic time-scale~(\ref{eq:tsto}), $t_{\rm sto}\simeq 86\myr$, is similar to the local dynamical time of the host galaxy, $\Omega_g^{-1}(20\kpc)\simeq 82\myr$. 

Left panel of Fig.~\ref{fig:a_2} shows that binary stars with a semimajor axis larger than the tidal radius~(\ref{eq:rt}), $a\gtrsim r_t(20\kpc)\simeq 0.98\kpc$, are progressively wiped out by the smooth tidal field, which leads to a semimajor axis distribution sharply truncated at $a_{\rm peak}\approx r_t$. Recall that the most-likely semimajor axis of pairs formed in isolation roughly corresponds to the average separation between stream particles, $a_{\rm peak}^{\rm isol}\sim D \simeq 4.3\kpc$. The fact that the tidal radius is much smaller than the mean interstellar distance in the tails, $r_t\ll D$, means that the statistical model presented in \S\ref{sec:model} largely overestimates the number of wide binaries that would survive to the present day.

The disruption of wide binaries is greatly enhanced in a clumpy tidal field. As expected, right panel of Fig.~\ref{fig:a_2} shows that differences between smooth and clumpy models arise on time-scales, $t\gtrsim t_{\rm sto}(20\kpc)\approx 86\myr$. On time-scales $t\gg t_{\rm sto}$ models that are acted on by tidal evaporation exhibit scale-free behaviours at large and small separations: (i) at $a\ll a_{\rm peak}$ the distribution preserves its original form, $p(a)\sim a^{1/2}$, expected from chance superposition of orbits in tidal streams (see Fig.~\ref{fig:bin_2}), whereas (ii) at large semimajor axes $a\gg a_{\rm peak}$ the perturbed distribution roughly follows \"Opik (1924)'s law, $p(a)\sim a^{-1}$ (black dashed line). This scale-free law is considerably shallower than the steep truncation induced by the smooth tidal field. This is because the tail $p(a)\sim a^{-1}$ is populated by particles drifting towards the escape energy, $E\to 0$ (see P19a for illustration of the random-walk process underwent by tidally-heated particles), while in a smooth tidal field the energy distribution is truncated at $E_t\equiv -Gm_b/(2 r_t)$. The above results can be straightforwardly combined to measure the survival fraction of wide binaries at a fixed semimajor axis $f_s(a,t)\equiv p(a,t)/p(a,t=0)$ on time-scales $t\gg t_{\rm sto}$: (i) in a smooth potential $f_s=1$ for $a\lesssim r_t$, and $f_s=0$ for $a\gtrsim r_t$, (ii) in a clumpy medium $f_s=1$ for $a\lesssim a_{\rm peak}$, dropping as $f_s\sim a^{-3/2}$ for $a\gtrsim a_{\rm peak}$.



\begin{figure}
\begin{center}
\includegraphics[width=86mm]{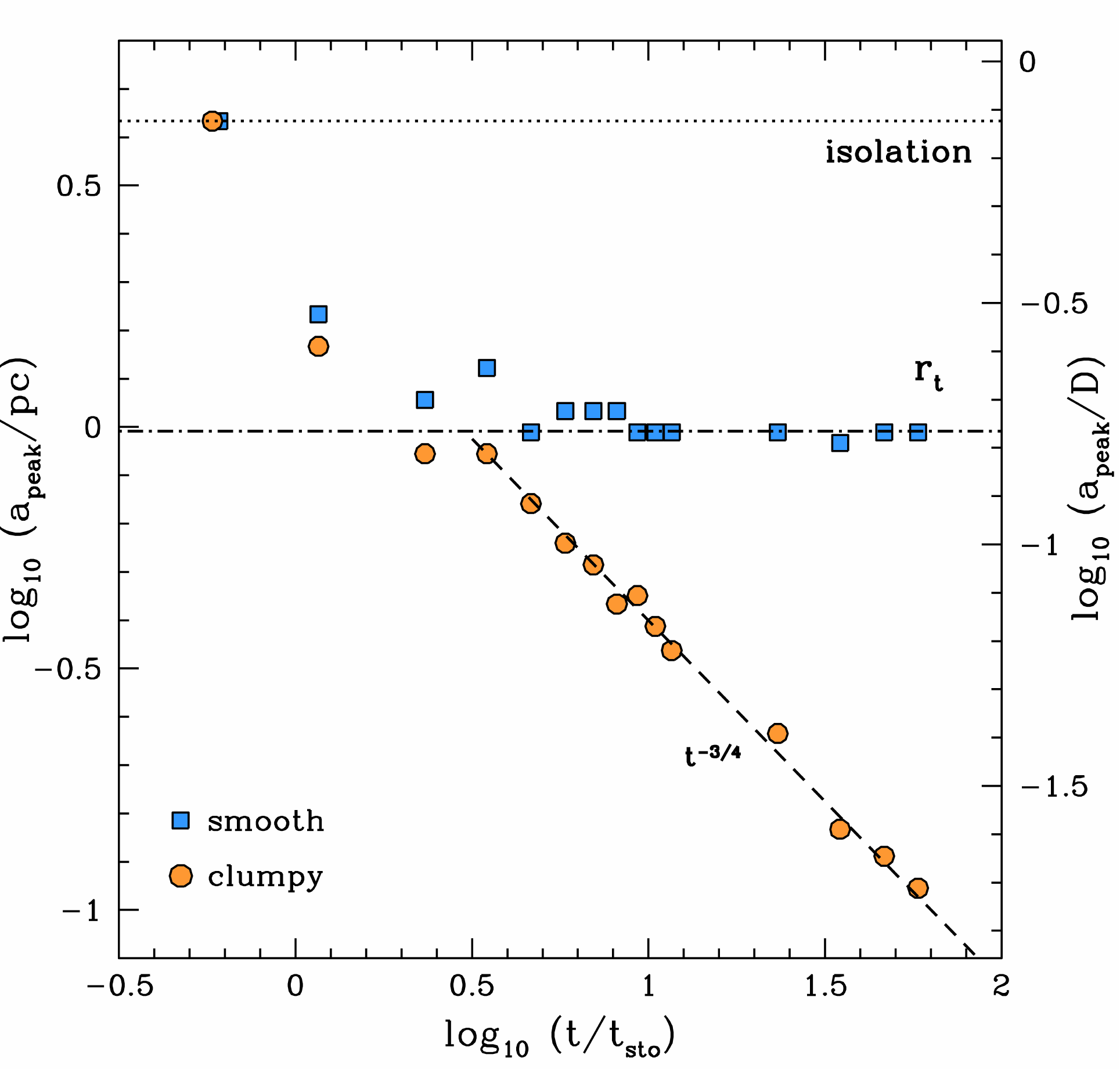}
\end{center}
\caption{Evolution of the peak semimajor axes ($a_{\rm peak}$) of the models shown in Fig.~\ref{fig:a_2} measured in parsecs (left vertical axis) and in units of the average stellar distance $(D)$ (right vertical axes), with time given in units of the stochastic time-scale, $t_{\rm sto}(20\kpc)=86\myr$. For reference, the value of $a_{\rm peak}$ found in isolation is shown with a dotted line. In a smooth host potential $a_{\rm peak}$ converges towards the tidal radius of the bound pair, $r_t$, given by Equation~(\ref{eq:rt}). In contrast, in a clumpy potential the peak semimajor axis decays monotonically as $a_{\rm peak}\sim (t/t_{\rm sto})^{-3/4}$ at $t\gg t_{\rm sto}$, with $t_{\rm sto}= 86\myr$ is given by Equation~(\ref{eq:tsto}).} 
\label{fig:apeak}
\end{figure}

Another remarkable effect of tidal evaporation is the progressive contraction of the peak semimajor axis as the length of the time interval increases. This is in stark contrast with the behaviour of $a_{\rm peak}$ in a smooth potential, which remains approximately constant at $a_{\rm peak}\approx r_t$. To inspect this result in detail, Fig.~\ref{fig:apeak} shows the location of $a_{\rm peak}$ as a function of time. As expected, differences begin smooth and clumpy models become noticeable on time-lengths comparable to the stochastic time-scale: (i) in a smooth potential, the value of $a_{\rm peak}$ quickly converges towards the local tidal radius (dotted-dashed line) given by Equation~(\ref{eq:rt}), (ii) in a clumpy potential it follows a scale-free behaviour $a_{\rm peak}\sim (t/t_{\rm sto})^{-3/4}$ (long-dashed line) at $t\gtrsim t_{\rm sto}$.

Measuring the peak semimajor axis in units of the average distance between stream stars (right vertical axis) shows that binaries acted on by a tidal field (either smooth or clumpy) have $a_{\rm max}\ll D$. In contrast, bound pairs formed in isolation show $a_{\rm max}\approx D$, which suggests that the ratio $a_{\rm max}/D$ can be used to gauge the level of disruption experienced by ultra-wide binaries in tidal streams.

\begin{figure}
\begin{center}
\includegraphics[width=86mm]{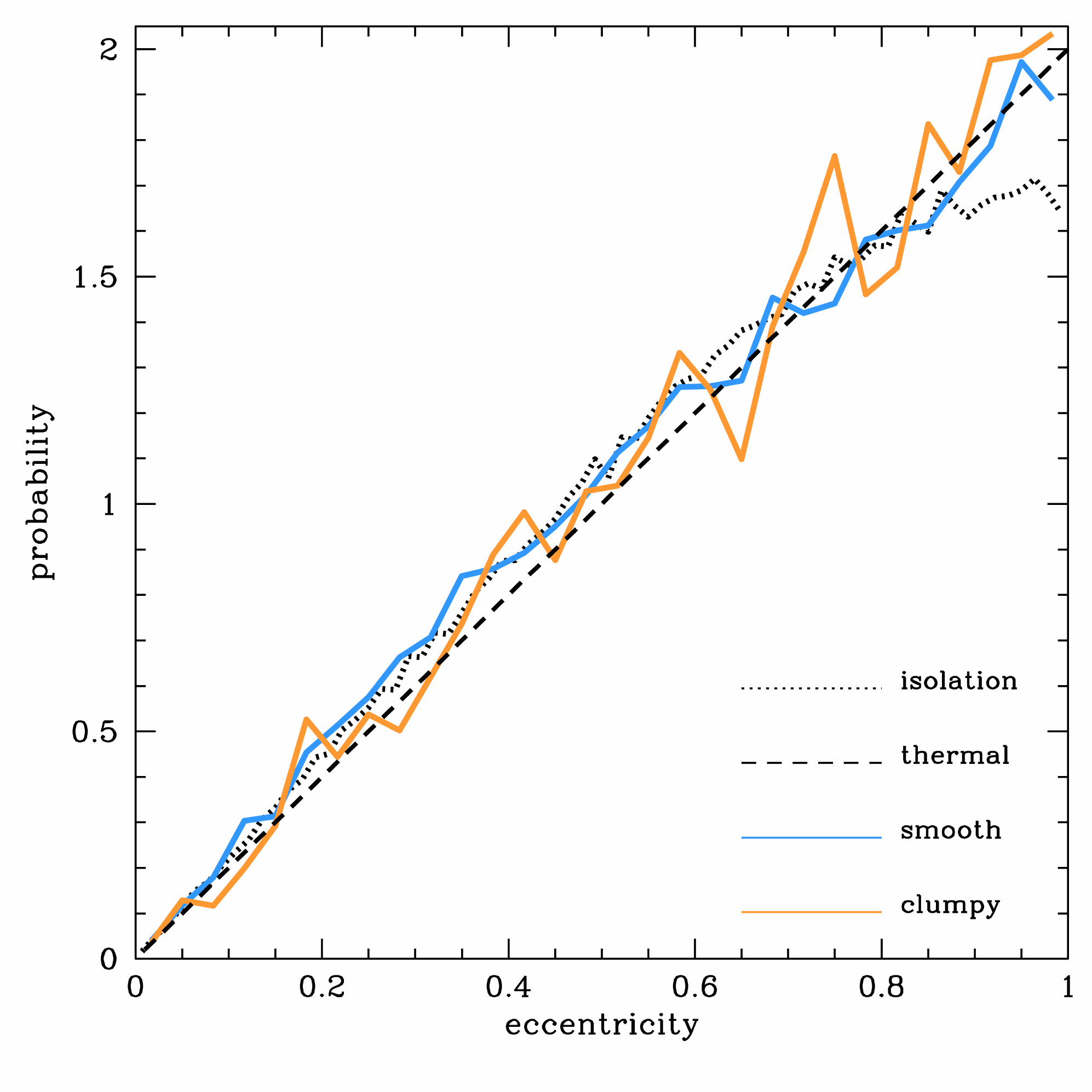}
\end{center}
\caption{Eccentricity distribution of binaries evolved $3\gyr$ in smooth and clumpy host potentials. For reference, the distribution found in isolation is shown with a black-dotted line, and the thermal function $p(e)\d e=2e\d e$ with a black-dashed line. } 
\label{fig:ecc_final}
\end{figure}

Finally, Fig.~\ref{fig:ecc_final} shows that the eccentricity distribution of binary stars evolving in smooth and clumpy potentials remains close to thermal at all times. 
We have explicitly checked that above results hold independently of the orbital radius of the progenitor cluster.

\section{Discussion}\label{sec:discuss}

\subsection{Limitations and follow-up work}\label{sec:limit}
Our analysis rests on a number of simplifying assumptions that are worth discussing here. The strongest limitation of this paper is the use of collision-less $N$-body models to follow the disruption of dense stellar clusters. Accounting for gravitational interactions between individual stars (within and without the cluster) will open up a number of interesting questions.

For example, for similar arguments as those laid out in \S\ref{sec:model}, binary and multiple-body ($N>2$) systems are expected to form within stellar clusters with high phase-space densities. The creation/destruction cycle of multiple stars inside a cluster is highly dynamical, as weakly-bound associations are created and disrupted on dynamical time-scale. Given that our $N$-body models do not capture this rich dynamical interplay, \S\ref{sec:nbody} estimates on the number of bound stellar pairs associated with the disruption of a single cluster are conservative, for they neglect the pre-existence of stellar associations that may end up in the material stripped from the progenitor cluster (K10), as well as soft binaries that form in the outskirts of star clusters expanding due to collisional relaxation (Moeckel \& Clarke 2011). According to K10, the existence of `primordial' binaries may facilitate the formation of wide triple and quadruple systems in tidal streams (see also Perets \& Kouwenhoven 2012).  
In addition, our $N$-body models do not contain a gaseous component, which may be an important ingredient to follow the formation of wide binaries during an early gas-expulsion phase (K10; Moeckel\& Bate 2010). 

The statistical model outlined in \S\ref{sec:model} describes the formation of wide binaries as a random superposition of stream orbits at a fixed snapshot. This picture neglects the self-gravity of individual stars and how it affects the relative trajectories of particle pairs as they move along the tidal tails. We have also ignored the formation of triple or higher-order multiple systems. Although the probability of an orbital {\it carambolage} involving several stream stars decreases with the multiplicity of the association, we did find such systems in the stream models of \S\ref{sec:nbody}. Given that multiplicity properties of ultra-wide binaries offer important tests on formation scenarios (e.g. Joncour et al. 2017), this issue is worth re-visiting in follow-up work.




In \S\ref{sec:nbody} we follow the evolution of clusters moving on circular disc orbits in the Galactic mean-field potential. This setup simplifies our analysis in two important ways. First, it allows us to model Galactic tidal field as a static function with no explicit time dependence. Yet, clusters typically move on eccentric orbits and experience a rapidly-varying tidal field and impulsive mass loss events at each pericentric passage. Numerical experiments not shown here indicate that most wide binaries form shortly after each pericentre, when the mass loss rate from the cluster peaks. Second, our $N$-body models also neglect orbital scattering between stream stars and halo substructures, which is expected to heat up the tidal tails (see P19b), thus making the coalescence of bound pairs less efficient with time.
In follow-up contributions we will extend our analysis to clusters on eccentric orbits and mimic the heating of tidal streams due to encounters with field substructures using the Monte-Carlo technique presented in P19b.

Section~\ref{sec:MC} analyzes the properties of self-gravitating binary systems with $E<0$. However, many ultra-wide binaries are continuously perturbed into and out of bound configurations. Because these stars have small relative velocities, one should expect long-range correlations in the positions and velocities of tidally-disrupted pairs for a time-scale comparable to the orbital period around the host (Jiang \& Tremaine 2010). Future modelling of wide binaries in tidal streams will incorporate a population of dissolved ($E>0$) binaries in the theoretical predictions, and not simply extrapolate the distributions plotted in Fig.~\ref{fig:a_2} at arbitrarily long distances.

Our analysis rests on the tidal approximation, which breaks down for binary separations larger than the mean distance between substructures, $a\gtrsim D$. Setting $a=D$ at a fixed clump density $\rho=M\,n$ defines a critical mass $M_{\rm crit}=2\pi \rho D^3$, such that the approximation holds for clump masses above $M\gtrsim M_{\rm crit}$. Binary stars moving in a medium of lighter point-mass objects, $M\lesssim M_{\rm crit}$ suffer `penetrating' encounters, which require solutions to two coupled differential equations, one for each individual binary member, with stochastic forces that can be assumed to be spatially uncorrelated (see P19b).

\subsection{Ultra-wide binaries in the field }
Section~\ref{sec:results} shows that the initial mass and size of a stellar cluster determines the number of bound pairs that can potentially form in its tidal tails.
 Given that the initial size of a cluster is tightly correlated with its mass, what systems are expected to dominate the population of ultra-wide binaries released into the field?

To answer this question it is useful to introduce a size function $r_c(M_c)$, which determines the variation of the scale-radius as a function of cluster mass, $M_c$. Following Choksi, Gnedin \& Li (2018), let us adopt a power-law relation
\begin{eqnarray}\label{eq:rcm}
r_c(M_c)=r_{c,0}\bigg(\frac{M_c}{M_0}\bigg)^\beta,
\end{eqnarray}
with $r_{c,0}=2.4\pc$, $M_0=2\times 10^5M_\odot$ and $\beta=1/3$. In addition, the initial cluster mass function observed in young massive star clusters in the Local Group and beyond (e.g. Portegies Zwart et al. 2010, Krumholz et al. 2019) can be fitted by another power law
\begin{eqnarray}\label{eq:nm}
\frac{\d N_c}{\d M_c}=N_{c,0}\bigg(\frac{M_c}{M_0}\bigg)^\alpha,
\end{eqnarray}
where $N_{c,0}$ is a normalization constant, and $\alpha=-2$.
For simplicity, Equation~(\ref{eq:nm}) ignores the exponential cutoff of the cluster mass function at the high mass end (e.g. Gieles et al. 2006), and simply assumes that the scale-free relation holds within a mass range $M_c\in(M_{c,1},M_{c,2})$, with $M_{c,2}\ggg M_{c,1}$.

The power-laws~(\ref{eq:rcm}) and~(\ref{eq:nm}) can be combined into a probability function (e.g. Pe\~narrubia 2018)
\begin{eqnarray}\label{eq:nmc}
\frac{\d^2 n_c}{\d M\d r_c}=\frac{n_c(R)}{M_0}\bigg(\frac{M_c}{M_0}\bigg)^\alpha\delta[r_c-r_c(M_c)],
\end{eqnarray}
where $\delta$ denotes Dirac's delta function, and $n_c(R)$ is the number density profile of clusters in the host galaxy.

Let us now assume that ultra-wide binaries form in the tidal debris of clusters that were disrupted a long time ago, $t_{\rm dis}\ll t_{\rm now}$.
Multiplying~(\ref{eq:Nbfinal}) and~(\ref{eq:nmc}), and integrating over mass and size yields the total number of ultra-wide binaries released into the field per unit volume
\begin{align}\label{eq:nbfield}
  n_b(R)&=\int_{M_{c,1}}^{M_{c,2}}\d M_c \int\d r_c \,\frac{\d^2 n_c}{\d M\d r_c}\,N_b(M_c,r_c)\\\nonumber
   &\approx 39.2\,n_c(R)\bigg(\frac{a}{r_{c,0}}\bigg)^{3/2}\bigg(\frac{M_0}{m_\star}\bigg)^{1/2} \\ \nonumber
 &\times\frac{1}{3/2+\alpha-3\beta/2} \bigg[\bigg(\frac{M_{c,2}}{M_0}\bigg)^{3/2+\alpha-3\beta/2}-\bigg(\frac{M_{c,1}}{M_0}\bigg)^{3/2+\alpha-3\beta/2}\bigg].
\end{align}
Clearly, the sign of the power-index of the right-hand term of~(\ref{eq:nbfield}) determines whether the formation of ultra-wide binaries is dominated by massive or low-mass clusters.
Inserting $\alpha=-2$ and $\beta=1/3$ yields $3/2+\alpha-3\beta/2=-1$, which indicates that the formation of ultra-wide binaries mainly occurs in the debris of low-mass stellar clusters, and that the total number of ultra-wide binaries released into the field is governed by the low-mass end of the initial cluster mass function\footnote{Recently, Choksi \& Kruijssen (2020) argue that clusters form with an initial power-law size function with an index $\beta=1/2$. Our conclusions do not change insofar as $3/2+\alpha-3\beta/2<0$. For $\alpha=-2$ this means $\beta >-1/3$.}. Indeed, setting $m_\star=0.5M_\odot$, $r_{c,0}=2.4\pc$, $M_0=2\times 10^5M_\odot$, $\alpha=-2$ and $\beta=1/3$ in Equation~(\ref{eq:nbfield}) returns a number density of ultra-wide binaries
\begin{align}\label{eq:nbfield2}
  n_b(R)\approx 1.1\times10^4\,n_c(R)\bigg(\frac{a}{2.4\pc}\bigg)^{3/2}\bigg(\frac{2\times10^5M_\odot}{M_{c,1}}\bigg),
\end{align}
which is inversely proportional to the minimum cluster mass of the initial mass function~(\ref{eq:nm}). According to Oey et al. (2004), the low-mass end  may be as small as a few solar masses, $M_{c,1}\sim M_\odot$, suggesting that the number density of ultra-wide binaries released into the field is several orders of magnitude above that of stellar clusters. However, it is worth bearing in mind the results of \S\ref{sec:disrup}, which show that the number of ultra-wide binaries that survive to the present day is but a small fraction of those that form, and that the precise number will depend on their formation time, their orbits in the mean-field potential, as well as on the clumpiness of the host galaxy.



\subsection{Ultra-wide binaries in cold tidal streams}
When modelling the dynamical evolution of bound pairs in the field, one of the chief unknowns is the formation time of these systems, which can vary substantially depending on whether binaries are found in star-forming regions of the Milky Way disc, or in the stellar halo.
This hindrance may be greatly alleviated by modelling the formation \& dissolution of ultra-wide binaries in the tidal tails of individual globular clusters. To date, there is about a dozen of known cold streams in the inner regions of the Galaxy (Malhan et al. 2018; Ibata et al. 2019). Among those, the best studied correspond to the stream associated with the globular cluster Palomar 5 (Odenkirchen et al. 2003), and GD-1 (Grillmair \& Dionatos 2006), whose progenitor remains undetected to date, likely because it was completely disrupted (Malhan \& Ibata 2019; de Boer et al. 2020). Bearing in mind the limitations discussed in \S\ref{sec:limit}, it interesting to estimate the number of ultra-wide binaries that can potentially form in those systems.
\begin{itemize}
\item {\it Pal 5 stream}. Palomar 5 is one of the star clusters with the lowest-density in the Galactic halo and it is best known for its stellar stream, which spans over 20 degrees across the sky. Using direct $N$-body tools, Kuepper et al. (2015) find that the best-fit initial conditions of the stream associated with Pal 5 correspond to a cluster with an initial mass $M_c=2.2\times 10^4M_\odot$ and a half-light radius, $r_h=11\pc$. Adopting a cored Dehnen (1993) profile, this translates into a scale-radius $r_c=r_h/3.85\approx 2.85\pc$.
Pal 5 is on the brink of full disruption (Erkal et al. 2017), which means $t_{\rm dis}\approx t_{\rm now}$ in our models. From~(\ref{eq:Nbfinal}), it follows that the number of ultra-wide binaries that form with a semimajor axis $a<1\pc$ in the Pal 5 stream is $N_b(a<1\pc)\approx 855$.
\item {\it GD1}. The GD-1 stream extends over $\sim 100^\circ$ on the sky, with a progenitor that remains undetected and was probably fully disrupted in the past. Webb \& Bovy (2018) find that the best-fit progenitor for GD1 stream is a cluster with a mass $M_c=1.6\times 10^4M_\odot$ and half-light radius $r_h=20\pc$. The length of the stream is consistent with a dynamical age of between $2$--$3\gyr$. Setting $t_{\rm dis}\ll t_{\rm now}$ in~(\ref{eq:Nbfinal}), and inserting the initial mass and size of the progenitor cluster then yields $N_b(a<1\pc)\approx 590$.
\end{itemize}

The above estimates suggest that hundreds of ultra-wide binaries may have populated the tidal streams of Pal 5 and GD1 for several Gyr. Section~\ref{sec:disrup} shows that the dynamical evolution of these objects depends very strongly on whether dark matter behaves as a smooth field on $\pc$ scales, or is made if compact objects, which can be used to test the presence of dark and visible clumps the Milky Way halo (Bahcall et al 1985; Weinberg et al. 1987; Chanam\'e \& Gould 2004; Quinn et al. 2009). This will be done in a separate contribution.
It is also interesting to notice that (1) the coalescence of ultra-wide binaries is considerably more efficient in tidal streams than in the outskirts of the progenitor clusters, which typically contain 1 per cluster (Moeckel \& Clarke 2011), and (2) the formation of ultra-wide binaries is not limited to a relatively short expansionary phase following gas expulsion from a star-forming region (see K10), but it spans over the entire dynamical evolution of clusters undergoing mass stripping, and will continue long after the progenitor system has been fully disrupted.

Furthermore, gradients in the abundance pattern of the progenitor globular cluster may be reflected in the relative composition of bound pairs (e.g. Andrews et al. 2019). Based on the results of \S\ref{sec:age}, a distinct chemical composition may be particularly visible in binary systems that form at the latest stages of the cluster evolution, as these pairs cover the largest spread in unbinding times.


\section{Summary}\label{sec:summary}
This paper studies the formation of ultra-wide binaries in the tidal tails of stellar clusters and their subsequent evolution in a clumpy Galactic environment. Our findings can be summarized as follows
\begin{itemize}
  \item In the scenario proposed here ultra-wide binaries arise via chance entrapment of unrelated stars in the tidal streams of disrupting clusters.
\item The rate at which bound pairs are created is proportional to the {\it local} phase-space density of stream stars, $Q=n/\sigma^3$, where $n$ is the number density and $\sigma$ is the velocity dispersion of stars neighbouring the pair.
  \item The formation of wide binaries is not limited to an early evolutionary phase of a star-forming region, but it spans over the dynamical evolution of tidally-stripped clusters, and continues in the tidal debris long after the progenitor system has been fully disrupted.
\item Ultra-wide binaries formed in tidal streams follow a universal semimajor axis distribution $p(a)\d a\sim a^{1/2}\d a$ on scales $a\ll D$, where $D=(2\pi n)^{-1/3}$ is the mean distance between stream stars, and an eccentricity distribution that is close to thermal, $p(e)\d e=2 e\d e$.
\item The spread of unbinding times, $\tau\equiv |\Delta t_{\rm unb}|$, can be calculated from the autocorrelation of the fractional mass-loss rate of the progenitor cluster $p(\tau)=\int_{-\infty}^{+\infty}\d t\,\mathcal{R}_{\rm unb}(t)\mathcal{R}_{\rm unb}(\tau+t)$, which scales as $p(\tau)\sim \tau^{-1.2}$ for $\tau\ll t_{\rm dis}$, where $t_{\rm dis}$ is the disruption time of the cluster. This implies that the coalescence of bound pairs is most likely to happen shortly after particles are stripped from the progenitor.
\item Most bound pairs formed in tidal streams are extremely fragile objects that can be disrupted by the smooth tidal field of the host as well as by passing substructures. The cumulative effect of random tidal fluctuations causes a progressive unbinding of self-gravitating objects known as ``tidal evaporation'', which dominates the removal of bound stellar pairs on a time-scale $t\gg t_{\rm sto}\sim \rho^{-1}$, where $\rho$ is the local density of compact substructures.
\item Disruption by the smooth field leads to a sharp truncation at the tidal radius, $p(a)\sim 0$ at $a\gg r_t$. In contrast, tidal evaporation leads to a semimajor distribution known as \"Opik (1924)'s law, $p(a)\sim a^{-1}$ at $a\gg a_{\rm peak}$, typically observed in young clusters and nearby OB associations (Kouwenhoven et al. 2007; Kraus \& Hillenbrand 2008, 2009). The peak semimajor axis contracts with time as $a_{\rm peak}\sim t^{-3/4}$ on a time-scale $t\gg t_{\rm sto}$, which implies that the population of ultra-wide binaries in tidal streams is progressively wiped out by a fluctuating tidal field. Interestingly, the semimajor axis distribution becomes {\it steeper} as the amount of substructures {\it decreases}, suggesting that the shallower separation function observed in the disc relative to that in the stellar halo (Tian et al. 2020) may be caused by tidal evaporation.
 \item The surviving population of bound pairs has a thermalized eccentricity distribution at all times, which agrees with observations of the population of wide binaries in the field (Tokovini 2020).
 \item The number of ultra-wide binaries released into the field is dominated by the disruption of low-mass stellar clusters.
 \item Hundreds of ultra-wide binaries are expected to populate the streams of individual globular clusters for several $\gyr$ in the stellar halo. The survival of weakly-bound binaries strongly depends on whether the halo potential behaves as a field or is made of compact objects, offering accurate targets to measure the clumpiness of the Galactic halo on $\sim\pc$ scales.
  \end{itemize}
As a final remark, it is worth noting that if the progenitor cluster contains Black Holes (BHs), the mechanism described above also predicts the formation of stellar-BH and BH-BH binaries in the associated tidal streams. This opens up a number of interesting questions regarding the stability \& detection of these systems (see Michaely \& Perets 2019), which will be explored in separate contributions.

\section{Acknowledgements}
It is a pleasure to thank Mark Gieles, Julio Chanam\'e and the anonymous referee for very useful comments.
\\\\
{\bf Data availability} The data underlying this article may be shared on reasonable request to the corresponding author.

\appendix
\begin{figure*}
\begin{center}
\includegraphics[width=160mm]{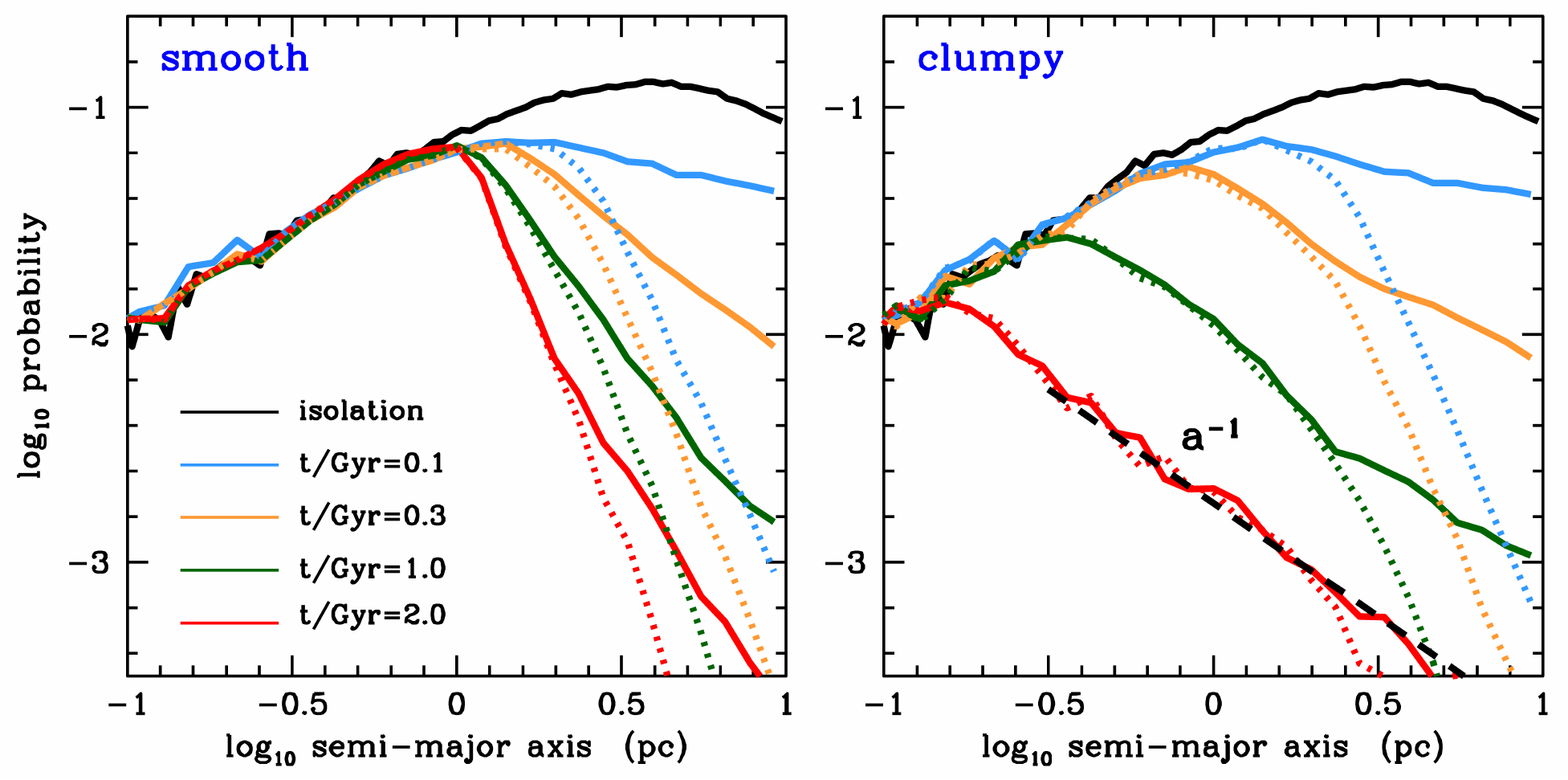}
\end{center}
\caption{Distribution of semimajor axes of wide binaries shown in Fig.~\ref{fig:a_2} with solid lines denoting pairs with self-gravitating energies $E<0$, and dotted lines showing particles with Jacobi energies $E_J<0$. Note the covergence of populations as {\it transient} pairs with $E<0$ and $E_J>0$ are disrupted by the external tidal field.} 
\label{fig:a_2_j}
\end{figure*}
\section{Jacobi energy}
This paper defines ``binary stars'' as particle pairs whose self-gravity is negative, $E<0$, thus ignoring the presence of an external host potential. Jiang \& Tremaine (2010) use a different definition that relies on the Jacobi energy $E_J=E+\Phi_c<0$, where $\Phi_c(r)=(1/2) \Omega_g^2 r^2$ is the centrifugal potential.
There are clear pros and cons in both definitions. E.g. in the absence of substructures, the Jacobi energy is an integral of motion, whereas the binding energy $E$ oscillates along the orbital phase. On the other hand, the Jacobi energy only exists for particle pairs on circular orbits around the host galaxy, which severely limits its applications. There are further reasons for inspecting self-gravitating energies rather than Jacobi integrals: (i) by ignoring the external potential, the statistical analysis of random pairs presented in \S2 becomes independent of the orbit/location of the pair in the host, (ii) self-gravitating energies can be computed directly from observations of the relative positions and velocities between stellar pairs with a given mass without making assumptions about the underlying host potential; and (iii) the impact of an external potential (smooth or clumpy) can be studied separately from the  binary formation process(es), which greatly simplifies our theoretical analysis.

It is worth stressing that the main results of the paper do not change significantly if we re-define ``bound pairs'' as particles with negative Jacobi energies, $E_J<0$. To illustrate this point, Fig.~\ref{fig:a_2_j} plots the semi-major axis distribution of particle pairs with with $E_J<0$ (dotted lines) and $E<0$ (solid lines). As expected from Fig.~\ref{fig:phase}, there is a large number of {\it transient} pairs that form with $E<0$ and $E_J>0$. Our analysis indicates that these objects are very weakly bound and do not last long as self-gravitating pairs. This is shown in the left panel of this figure, which shows that most transient pairs are destroyed by the smooth tidal field within one dynamical time $\Omega_g^{-1} \simeq 100 \myr$. As time goes by, solid and dotted lines converge to each other. The convergence is much faster in clumpy potentials, where the disruption of transient pairs is more efficient.

\end{document}